\documentclass[structabstract]{aa}  
\usepackage{xspace}
\usepackage{epsfig}
\usepackage{epstopdf}
\usepackage{graphicx}
\usepackage{natbib}
\usepackage{subfigure}
\usepackage{color}

\newcommand{\micron}{\ensuremath{\mu\mathrm{m}}\xspace}
\newcommand{\Vtot}{\ensuremath{\widetilde{V}_{\mathrm{tot}}}\xspace}

\newcommand{\BoL}{\bigg(\frac{\mathbf{b}}{\lambda}\bigg)}
\newcommand{\BoLL}{\bigg(\frac{\mathbf{b}}{\lambda}, \lambda \bigg)}
\newcommand{\BoLs}{\big(\frac{\mathbf{b}}{\lambda}\big)}
\newcommand{\denv}{\ensuremath{d_{\mathrm{env}}}\xspace}
\newcommand{\Venv}{\ensuremath{\widetilde{V}_{\mathrm{env}}}\xspace}
\newcommand{\fso}{\ensuremath{f_*^0}\xspace}
\newcommand{\uv}{\ensuremath{(u, v)}}
\newcommand{\modif}[1]{#1}

\usepackage{txfonts}
\begin{document}
   \title{SPARCO : a semi-parametric approach for image reconstruction
     of chromatic objects} 
     \subtitle{ Application to young stellar objects}
   \author{J. Kluska  \inst{1}
          \and
          F. Malbet \inst{1}
          \and
          J.-P. Berger \inst{2}
          \and
          F. Baron \inst{3}
          \and
          B. Lazareff \inst{1}
           \and
          J.-B. Le Bouquin \inst{1}
           \and
          J. D. Monnier \inst{4}
          \and
          F. Soulez \inst{5}
          \and
          E. Thi\'ebaut \inst{5}
          }
         
   \institute{Institut de Plan\'etologie et d'Astrophysique de Grenoble, UJF, CNRS,
              414 rue de la piscine, 38400 Saint Martin d'H\`eres, France\\
              \email{jacques.kluska@obs.ujf-grenoble.fr}
         \and
            European Southern Observatory, Alonso de Cord\'ova 3107, Vitacura, Santiago, Chile
            \and
        	   Center for High Angular Resolution Astronomy, Georgia State University, PO Box 3969, Atlanta, GA 30302, USA
	    \and
	    University of Michigan Astronomy Department, 941 Dennison Bldg, Ann Arbor, MI 48109-1090, USA
         \and
        	   CRAL, Observatoire de Lyon, CNRS, Univ. Lyon 1, Ecole Normale Sup\'erieure de Lyon (France)
            }
     
\date{Received 28 October 2013, accepted 11 March 2014}

  \abstract
  {
    The emergence of optical interferometers with three and more telescopes allows image reconstruction of astronomical objects at the milliarcsecond scale. However, some objects contain components with very different spectral energy distributions (SED; i.e. different temperatures), which produces strong chromatic effects on the interferograms that have to be managed with care by image reconstruction algorithms. For example, the gray approximation for the image reconstruction process \modif{results in a degraded} image if the total \uv-coverage given by the spectral supersynthesis is used.
}
  {The relative flux contribution of the central object and an extended structure changes with wavelength for different temperatures. For young stellar objects, the known characteristics of the central object (i.e., stellar SED), or even the fit of the spectral index and the relative flux ratio, can be used to model the central star while reconstructing the image of the extended structure separately.}
  {We present a new method, called \texttt{SPARCO} (semi-parametric algorithm for the image reconstruction of chromatic objects), which describes the spectral characteristics \modif{of both the central object and the extended structure} to consider them properly when reconstructing the image of the surrounding environment. We adapted two image-reconstruction codes (\texttt{Macim}, \texttt{Squeeze}, and \texttt{MiRA}) to implement this new prescription.}
  {\texttt{SPARCO} is applied using \texttt{Macim}, \texttt{Squeeze} and \texttt{MiRA} on a young stellar object model and also on literature data on HR 5999 in the near-infrared with the VLTI. We obtain smoother images of the modeled circumstellar emission and improve the $\chi^2$ \modif{by} a factor \modif{9}. 
  }
  {This method paves the way to improved aperture-synthesis imaging of several young stellar objects with existing datasets. More generally, the approach can be used on astrophysical sources with similar features such as active galactic nuclei, planetary nebulae, and asymptotic giant branch stars.} 
\maketitle

%
\section{Introduction}

The number of aperture-synthesis images based on optical long-baseline
interferometry measurements has recently increased thanks to easier access to
visible and infrared interferometers. The interferometry technique has now
reached a technical maturity level that opens new avenues for numerous
astrophysical topics that require milliarcsecond model-independent imaging
\citep{2012A&ARv..20...53B}. Image reconstruction \citep[see review][]{2013EAS....59..157T} is
the key to achieve the most probable, noncommittal images following some global
constraints (image positivity, size of the
support, regularization, etc.). A thorough study
\citep{Renard} has shown the limitations of image reconstruction, but never challenged the type of regularizations in use. The first images of the inner
regions of the environment have been obtained around the young
stellar objects HD\,163296 \citep{HD163296} and HR\,5999 \citep{HR5999}, revealing the structure around the central
objects. However, one main caveat of the images reconstructed from spectrally dispersed instruments is that the visibilities
measured at different wavelengths have been assumed to come from a \emph{gray}
object. 
In addition, the central star has a much higher
surface brightness than the surrounding emission, a contrast problem that limits the reliability and the quality of image reconstruction and complicates serious analysis of the morphology of the circumstellar material. These concerns motivate one to reconstruct an image of the envelope alone \modif{without the star in the image}, an approach considered first in self-aperture masking techniques \citep{2004ApJ...605..436M}. 

Some of the interferometric visibilities obtained with spectral
resolution on young stellar objects (YSO), mainly in the near-infrared (NIR), have
been seriously affected by strong spectral dependence. For example, the visibilities of the object MWC\,158, as measured with VLTI/AMBER
in the $K$ band and in the $H$ band are spread over a broad range \citep[see Fig.~5 of][]{2011AA...528A..20B}. 
First believed to be \emph{quality (...) clearly lower compared to the K-band data} \citep{2011AA...528A..20B}, the increase of
the visibilities of the $H$-band data has been observed not only on VLTI/AMBER, but also
with VLTI/PIONIER. This indication caused us to consider
an astrophysical interpretation: the change of visibility is generated by the
different chromatic behavior between the central star that peaks in the visible and the
circumstellar material that radiates out mainly in the near-infrared \citep{2012SPIE.8445E..0OK}. 

The star contribution is highest at the shortest wavelengths and becomes moderate at longer wavelengths. 
Classic gray algorithms assume the same brightness distribution for all wavelengths.
Consequently, they cannot 
satisfactorily reconstruct an image of a chromatic object with spectral supersynthesis. 
We need to include the chromatic effect induced by the physical properties of the target
in the image processing to retrieve a good intensity map of the observed target.

In this paper, we present a semi-parametric approach that includes the knowledge 
of the \modif{relative stellar and environment spectral} properties in optical interferometry image reconstruction. As previously demonstrated in parametric modeling of optical interferometric data (e.g. \citet{2012ApJ...752...11K} ), we can
directly take the object chromaticity into account in the process of image reconstruction, which improves the final fit.

This approach is called SPARCO (semi-parametric approach for image reconstruction of chromatic objects) and consists of separating a well-known object (e.g. the central star of a YSO) and its complex, unknown environment (e.g. its dusty disk). The star is modeled by a parametric model (that can include hydrostatic models or binaries) and the environment by the reconstructed image. The chromatism is reproduced by changing the flux ratio between the two components across the observed bandwidth.

To present the methods, we focus on YSOs because the star can be modeled at first order in the NIR by an unresolved component and the flux ratio \modif{can be represented by a power law with a good approximation}. Moreover, the environment is poorly known and is complex. The application of this technique to this type of objects is therefore important.

However, this method does not apply only to YSO,
but can also be used in any system where a known source is present that
displays a spectral behavior different from the rest of the emitting material in the optical. For instance, the accretion disk of an active galactic nuclei is considered as unresolved in order to retrieve its environment \citep{2013ApJ...775L..36K}. A method of separation of the star from its environment was invoked for asymptotic giant branch stars \citep{2013A&A...559A.111H}, or planetary nebulae \citep{2006A&A...448..203L}.

In Sect.~\ref{sec:sep} we demonstrate that the \texttt{SPARCO} approach allows proper modeling of the interferometric observables of  YSOs, especially their chromatic content. We show how standard image reconstruction algorithms can be modified accordingly. In Sect.~\ref{sec:imgrec} we validate the method on the
model of a realistic YSO. 
In Sect.~\ref{sec:dis}, we further discuss important aspects of the method in detail. 
We finish in Sect.~\ref{sec:HR5999} by
applying \texttt{SPARCO} on actual data used by \citet{HR5999}
to reveal the circumstellar environment of HR\,5999, and we compare it with a
previous analysis.

%

\section{\texttt{SPARCO} method}
\label{sec:sep}

\subsection{Fluxes from the star and the environment}
\label{sec:chrom}

\begin{figure*}[!t]
 \centering
 \includegraphics[width=8cm]{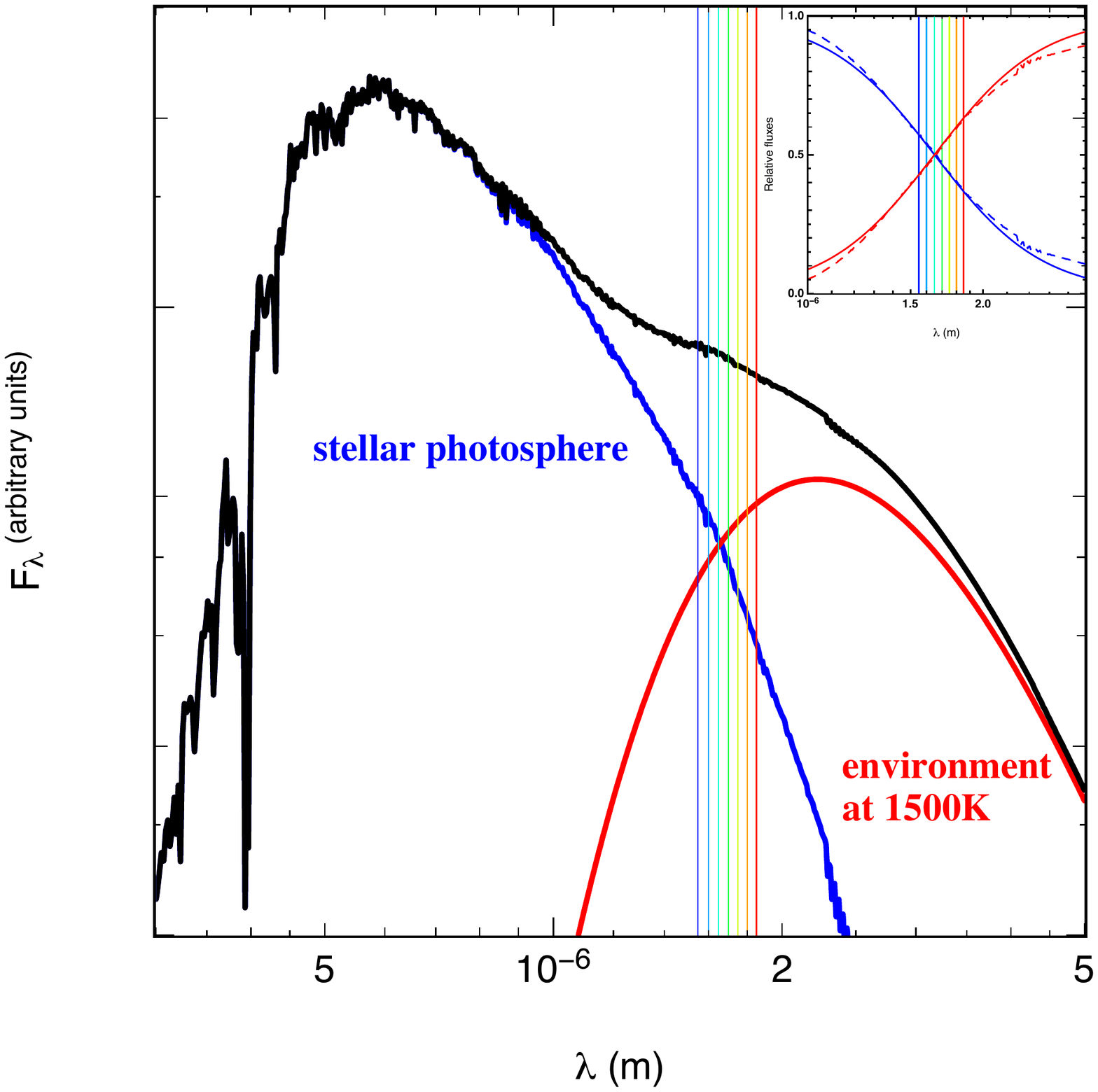} 
 \includegraphics[width=5cm]{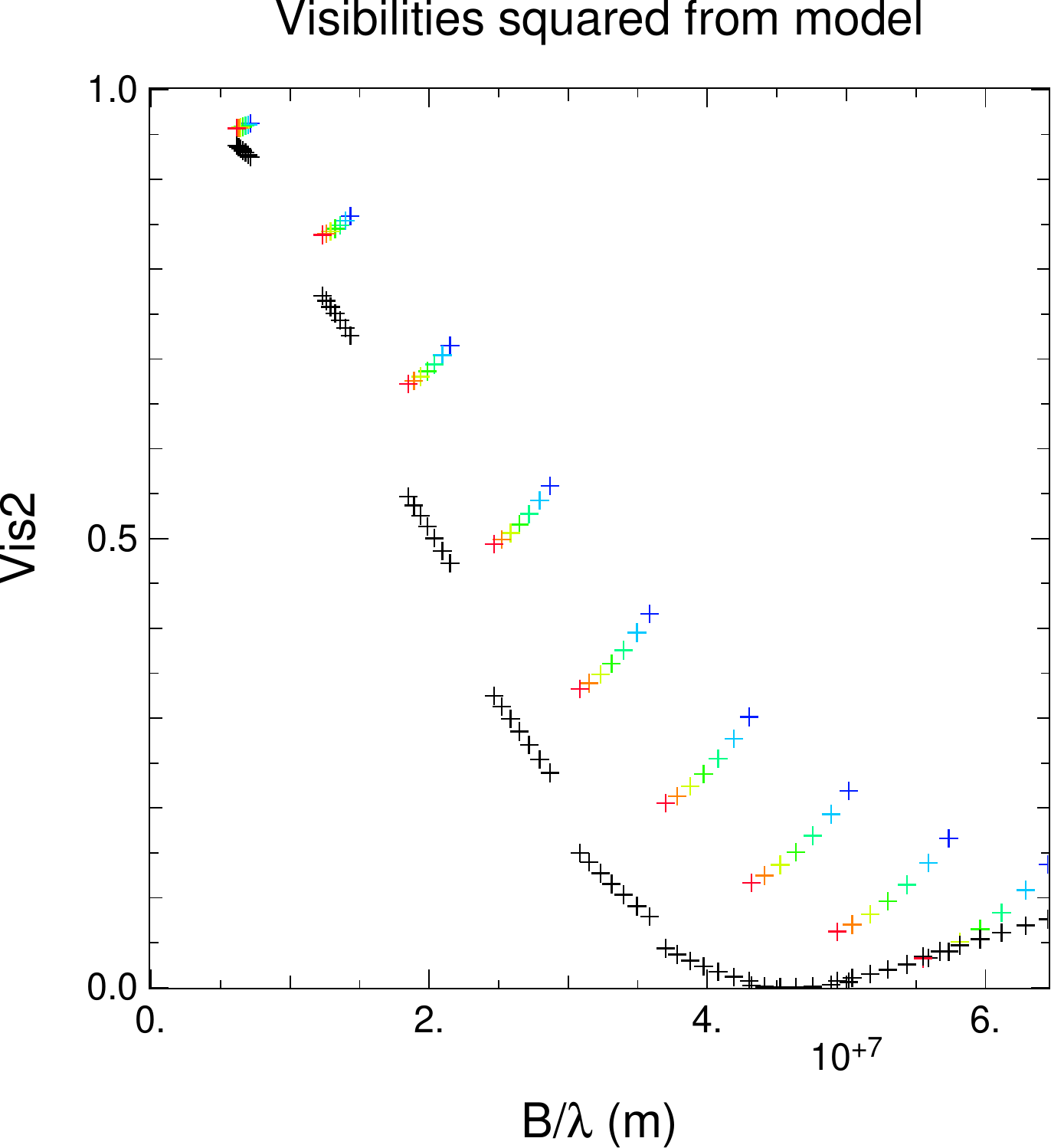}
 \caption{Left: SED of a young stellar object. In blue:
   the stellar photosphere (represented as a black body) at $10\,000\,\mathrm{K}$; in red: the
   environment black body at $1500\,\mathrm{K}$; and in black: the sum of the two
   contributions. The vertical color lines are the spectral channels of
   PIONIER. The two contributions cross in the $H$
   band. In the top right corner as a dashed line: the flux ratios of every component from the SED, 
   as a full line: the result of modeling the fluxes with power laws. Right: in black, expected visibilities from the extended structure
   alone. In color: the star contribution has been added and colors follow the
   standard color code (blue: shortest wavelength, red: longest wavelength). The visibilities increase
   with shorter wavelengths because the stellar contribution is higher.}
 \label{fig:SED}
\end{figure*}
 
A YSO, consisting of a star and a dusty environment, has  two main components in the spectral
energy distribution\footnote{The presence of an additional hot inner component as suggested by \citet{2008ApJ...677L..51T} and \citet{HR5999} is discussed Sect.~\ref{sec:HR5999}}. 
The photosphere dominates from the UV domain to the visible.
The contribution of the environment, which is mainly dusty and
occurs at lower temperature ($T<1500\,\mathrm{K}$), prevails from the infrared to 
radio wavelengths (see Fig.~\ref{fig:SED} Left).

For example, the effective temperature of Herbig Ae/Be stars is
$T\approx10\,000\,\mathrm{K}$. The spectral domain where the total flux is no
longer dominated by the stellar flux \modif{is} in the NIR where the
emission of the environment increases steeply. Typically, the stellar SED is
in the Rayleigh-Jeans regime in the $H$ band
($1.65\,\micron$), while the environment is in the Wien
regime. In these spectral bands ($J$,$H$ and $K$ bands for young stellar objects), the shortest wavelengths are dominated by the
photosphere flux, the longest ones by the environment flux.
These two components (the star and the dusty environment) have different spatial
extents, which can be resolved by current optical interferometers.

Some interferometric instruments (AMBER, PIONIER on the VLTI, and MIRC on the CHARA interferometer)
cover the $H$ band with several spectral channels. The change of flux ratio (star/environement) for each channel following the
physical laws described above implies a strong chromatic effect
in the visibilities called star/environment chromatism (see Fig.~\ref{fig:SED} Left).

When considering low spectral resolution within a single near-IR band, this effect can be described by the following two parameters:
\begin{itemize}
\item $f_*^0$: \textit{The stellar-to-total-flux ratio at wavelength $\lambda_0$.} This parameter sets the flux balance between the two components at the reference wavelength $\lambda_0$. $\lambda_0$ is arbitrary chosen (for example, the central wavelength of observations).

\item  $\denv$: \textit{the spectral index for the circumstellar environment.} For interferometric
data, only the difference between the spectral indexes of the two components
matters to account for the chromatism. In the NIR, the star emission occurs in
the Rayleigh-Jeans regime and can be approximated by
$F^*_{\lambda} \propto \lambda^{-4}$. The parameter that sets the index
difference is the absolute spectral index of the dust,
$F^{env}_{\lambda} \propto \lambda^{d_{env}}$.
\end{itemize}
The total flux $f_{tot}$ at a wavelength $\lambda$ normalized by the total flux at $\lambda_{0}$ can therefore be written as 
\begin{equation}
f_{tot}(\lambda) = f_*^0 \Bigg(\frac{\lambda}{\lambda_0}\Bigg)^{-4} + (1 -
f_*^0) \Bigg(\frac{\lambda}{\lambda_0}\Bigg)^{\denv}.
\label{eqn:ftot}
\end{equation}

This model is sufficient to describe the continuum emission of the object in a NIR band (especially the $H$ band). With additional effort, complex flux distribution (e.g. adding lines) can be used to model observations at higher spectral resolution, following Eq.~(\ref{eqn:gen}).

\subsection{Visibilities from the star and the environment}

The total complex visibilities are the sum of the stellar visibilities
and visibilities from the environmental components weighted by
their fluxes at the corresponding wavelengths ($\lambda$),
\begin{equation}
\label{eqn:gen}
f_{\mathrm{tot}}(\lambda) \Vtot \BoLL = f_*(\lambda) \widetilde{V}_* \BoL + f_{\mathrm{env}} (\lambda) \Venv \BoL,
\end{equation}
with $f_{tot}(\lambda) $ given by Eq.~(\ref{eqn:ftot}), $f_*(\lambda) $ the stellar flux, $f_{env}(\lambda) $ the flux of the environment, $\widetilde{V}_*$ the stellar visibility, $\Venv$ the visibility of the environment. Finally, $\mathbf{b}$ is the interferometric baseline vector projected on-sky. The visibility depends on $\mathbf{b}/\lambda$, which is the spatial frequency. In this paper all the quantities with a tilde (as $\Venv$) are complex numbers.

In YSO, the apparent size of the central star can be considered as unresolved. For instance, a Herbig
AeBe star at a typical distance of $140\,\mathrm{pc}$ (distance of the
Taurus star-forming region) and with a radius of
$5\,\mathrm{R}_{\odot}$ has an angular radius of
$\approx0.17\,\mathrm{mas}$. For a $100\,\mathrm{m}$ baseline and for the $H$ band, this
corresponds to a visibility $V \approx 0.997$. Hereafter we model this star with an unresolved component with $V_* \approx 1$.
By developing the total flux term, we have
\begin{equation}
\Vtot \BoLL =\frac{f_*(\lambda) + f_{\mathrm{env}} (\lambda) \Venv \BoLs }{f_*(\lambda) + f_{\mathrm{env}} (\lambda)},
\label{eqn:chrom1}
\end{equation}
where $\Venv$ is the visibility of the environment alone, derived from its brightness distribution by a Fourier transform. With the two parameters
described in Sect.~\ref{sec:chrom} we can rewrite Eq.~(\ref{eqn:chrom1}) as follows:
\begin{equation}
\Vtot \BoLL =\frac{f_*^0 (\frac{\lambda}{\lambda_0})^{-4} + (1 - f_*^0) (\frac{\lambda}{\lambda_0})^{\denv} \Venv \BoLs }{f_*^0 (\frac{\lambda}{\lambda_0})^{-4} + (1 - f_*^0) (\frac{\lambda}{\lambda_0})^{\denv}}.
\label{eqn:chrom2}
\end{equation}

\subsection{Image reconstruction}
\label{sec:imgrec_base}

$\Venv$ is retrieved by making the Fourier transform of its image. This image is obtained by a well-known image-reconstruction process. The goal is to retrieve the most probable image given the dataset and some assumption called regularizers (e.g. image positivity). To reconstruct the image we have to solve an ill-posed inverse problem by minimizing the function $\mathcal{J}$ defined as \citep{ImgRec}
\begin{equation}
\mathcal{J}(\textbf{x}) = \mathcal{J}_{\mathrm{data}}(\textbf{x}) + \mu \mathcal{J}_{\mathrm{rgl}}(\textbf{x}),
\label{eqn:f}
\end{equation}
$\mathcal{J}$ being the global distance to minimize,
$\mathcal{J}_{\mathrm{data}}(\textbf{x})$ the distance to the data (reduced
$\chi^2$), $\mathcal{J}_{\mathrm{rgl}}(\textbf{x})$ the regularization distance,
$\textbf{x}$ the image pixel values, and $\mu$ the
regularization weight.  The choice of the hyper-parameter $\mu$ is discussed in Appendix~\ref{app:rgl}. 
For more information about the regularization see \cite{Renard}.

Various image reconstruction algorithms exist (e.g. \texttt{MiRA} \citep{MiRA}, \texttt{Macim} \citep{Macim}, and \texttt{Squeeze} \citep{Squeeze}). They mainly differ by the way they minimize the function $\mathcal{J}$ (gradient descent or Monte Carlo Markov chain (MCMC) minimization).

The image-reconstruction process included in \texttt{SPARCO} is monochromatic. In
other words, the object intensity distribution is
wavelength independent. 
This image changes its flux ratio across the observed band only thanks to Eq.~\ref{eqn:chrom1}. 


Because most algorithms compute the complex
visibilities of the image, our method can be implemented in any
of them. They have to be modified to include Eq.~(\ref{eqn:chrom2}) in their iterative computation of visibilities ($\Venv$ being the complex
visibilities of the image). 
If the image-reconstruction algorithm is
based on a gradient descend, the gradient needs to be multiplied by the
factor $\frac{ \partial \Vtot} {\partial \Venv}$ (which is
the environment-to-total-flux ratio).
Current MCMC reconstruction algorithms (i.e. \texttt{Macim} and \texttt{Squeeze}) use stochastic steps whose modifications to the $\chi^2$ are computed by finite difference, and thus do not require any gradient evaluation.

The proposed method couples (1) the fitting of a parametric model for one part of the object with (2) a simultaneous image reconstruction of the remaining part. The details of the algorithm are described in Appendix~\ref{app:fit}. 

\section{Numerical validation}
\label{sec:imgrec}
To validate our method and estimate its capability to retrieve object features, we built a synthetic model. This geometrical model of an unresolved star and its surrounding environment includes the chromatic effect: the star is hotter than its environment. This object was used to simulate interferometric observations in a realistic configuration.

\subsection{Model description}

\begin{figure*}[!t]
\centering
\includegraphics[width=15cm]{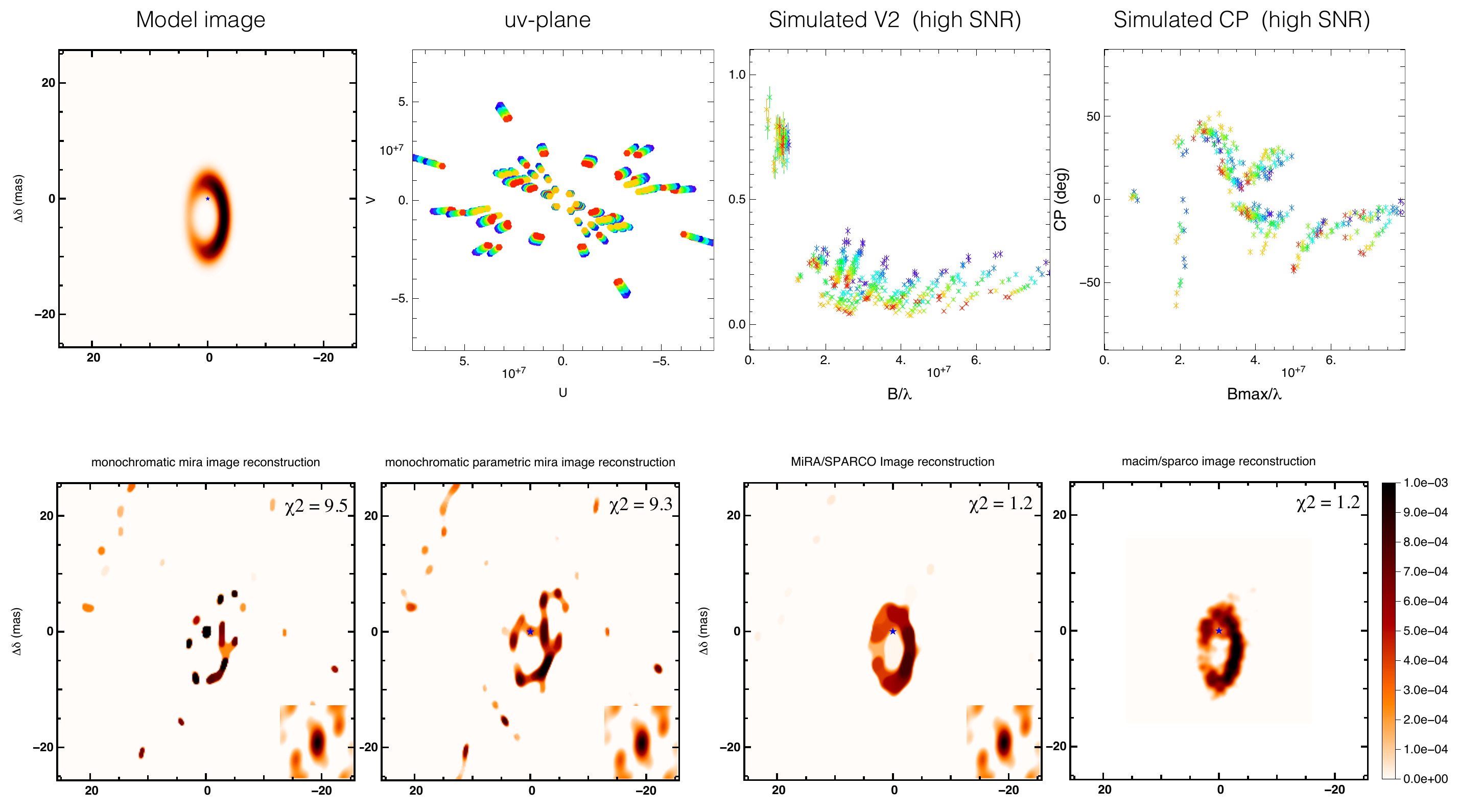}
\caption{Top: the image of the ring with the star represented in red (model), the \uv-plan used in the simulation, and the simulated dataset. Bottom: image reconstructions with the four cases discussed in Sect.~\ref{sec:cases}: 1) $\fso=0$ classical gray image reconstruction; 2) $\fso=0.4 and \denv=d_{\mathrm{star}}$; 3) $\fso = 0.4 and d_{\mathrm{star}}=-4 and \denv=1$ with MiRA/SPARCO; 4) idem with MACIM/SPARCO.}
\label{fig:img}
\end{figure*}

We used an analytical model consisting of an unresolved star and its environment. The star was assumed to be in the Rayleigh-Jeans regime so that its spectral dependence is: $F^*_{\lambda} \propto \lambda^{-4}$. Its spectral index is then $-4$. The star contributes 40\% of the flux at 1.65$\mu$m ($\fso=0.4$).

The environment was modeled by a Gaussian ring with a radius of 6 mas inclined by 60 degrees. An azimuthal modulation set the ring flux to vary as the cosine of the azimuthal angle. This reproduces the asymmetry generated by radiative transfer effects \citep{2006ApJ...647..444M}. The Gaussian thickness of the ring (the FWHM of the Gaussian with which we convolved the infinitesimal ring) is 2.4 mas. The center of the ring is shifted by 3 mas to the south with respect to the star. This shift can reproduce a perturbation induced by a companion or an inclination effect.  The environment spectral index is $\denv = 1$, which is the logarithmic black body-curve derivative at $\lambda_0$ and at a temperature of $\approx 1400\,\mathrm{K}$.  At 1.5$\mu$m the star carries $\sim55\%$ of the flux and $\sim38\%$ at 1.8$\mu$m. This chromatism is strong enough to significantly affect the data. All the parameters of this model are summarized in Table~\ref{tab:model}.

From this model we simulated a realistic dataset as it would have been obtained by the PIONIER instrument \citep{JBLB}. 
We used a \modif{actual} \uv-plan made with PIONIER (see Fig.~\ref{fig:img}) consisting of 14 pointings of four telescopes on three different configurations available at the VLTI (two pointings on small configuration, nine on the medium one, and three on the large one). One pointing represents 40 minutes of observation (calibrators included). Each point on each baseline is spectrally dispersed onto seven channels across the H band. PIONIER only provides $\mathrm{V}^2$ and closure phases (CP; no complex visibilities, no complex differential visibilities). We added realistic noise to the data by selecting three regimes: high, intermediate, and low flux (See Table~\ref{tab:err}).

The resulting dataset is shown in Fig.~\ref{fig:img}, top. This artificial dataset is qualitatively similar to real observations obtained with existing optical interferometers \citep[e.g.][]{2012SPIE.8445E..0OK}. The overall circumstellar structure is resolved by the longest baselines, and the V$^2$ \modif{data} display strong chromaticity effects.

\begin{table}
\caption{Model parameters}             
\label{tab:model}      
\centering                          
\begin{tabular}{c c c c}        
\hline\hline                 
Parameter & Value & Unit & Comments \\    
\hline                        
   $f_*^0$ & 0.4 & &  \\      
   $d_{\mathrm{env}}$ & 1.0 & &  \\
   $x_s$ & 0.0 & mas & star east shift    \\
   $y_s$ & 3.0 & mas  & star north shift \\
   $R$ & 6.0 & mas & ring radius   \\ 
   $i$ & 60 & deg & inclination \\
   $\theta$ & 0 & deg & P.A. of the major axis \\
   $w$ & 2.4 & mas & Gaussian width of the ring  \\
   $c1$ & 0.0 & & azim. mod. through major axis  \\
   $s1$ & -0.4 &  & azim. mod. through minor axis  \\
\hline                                   
\end{tabular}
\end{table}

\begin{table}
\caption{Model noise estimated from \citet{JBLB}.}             
\label{tab:err}      
\centering                          
\begin{tabular}{c c c c}        
\hline\hline                 
Noise & Hmag = 3 & Hmag = 5 & Hmag  = 7\\    
\hline                        
   $\sigma_{V^2}$ & 0.05 $V^2$ + 0.001 & 0.05 $V^2$ + 0.003 & 0.1 $V^2$ + 0.003 \\      
   $\sigma_{CP}$ & 2deg & 5deg & 10deg \\
\hline                                   
\end{tabular}
\end{table}

%
\subsection{Validation of the chromatic image-reconstruction method}
\label{sec:gray}

In this subsection we \modif{assume} a perfect knowledge of the object's chromatic parameters. We compare the images obtained with the classical "gray" approach to the \texttt{SPARCO} approach and check the validity in different signal-to-noise regimes.

Except when explicitly stated, the images were reconstructed with the {\tt MiRA} algorithm \modif{which} computes the image following a gradient method. The regularization used was "total variation", which was described in \citet{Renard} as the most successful regularization for astrophysical objects. The process of selecting the optical regularization weight ($\mu$, see Eq.~(\ref{eqn:f})) is described in Appendix~\ref{app:rgl}. In all cases, we performed the reconstruction with $\mu = 1500$. We emphasize that the regularization and its hyper-parameter has to be adapted to every reconstructed object. One has to be careful when choosing the value of the hyper-parameter $\mu$ although we show (in Appendix~\ref{app:rgl}) the low sensitivity of the method to the choice of regularization type. Appendix~\ref{app:rgl} also presents several methods that help to chose the regularization. We reconstructed images (see Fig.~\ref{fig:img}) with 256$\times$256 pixels of 0.2 mas each with three different methods:
\begin{enumerate}
\label{sec:cases}
\item A classic, gray image reconstruction, that is, with $\fso=0$ (in which case the $\denv$ parameter has no meaning).
\item An image reconstruction considering a central point source ($\fso=0.4$), but without taking into account the difference of spectral indicies ($\denv=d_{\mathrm{star}}=-4$).
\item An image reconstruction with the full SPARCO approach, that is, considering the difference of spectral index between the star and the environment ($\fso=0.4, d_{\mathrm{star}}=-4, \denv=1$). To show the effect of the SPARCO approach we implemented it in three different image-reconstruction algorithms (\texttt{MACIM}, \texttt{Squeeze} and \texttt{MiRA}).
\end{enumerate}

The reconstructed images are shown in Fig.~\ref{fig:img}, bottom. In the classical gray case the algorithm poorly fits the data ($\chi^2=9.3$). It is possible to lower the $\chi^2$ by \modif{significantly} increasing the field of view of the image and the number of pixels. \modif{With} these additional degrees of freedom, the algorithm is able to reproduce the chromatic effect by adding ripples in the Fourier space. This creates strong artifacts at both small and large separations in the image. Somehow it replaces chromatic effects by spatial artifacts linked with the Fourier sampling. For instance, the ring P.A. is related to the \uv-plan orientation. These effects are still observed when subtracting the star monochromatically.

The full \texttt{SPARCO} approach allows one to reach a good $\chi^2$ ($\approx1.2$). The reconstructed image does not show strong artifacts. 
The P.A. of the ring, its offset, and its azimuthal modulation are correctly reproduced. 
We conclude that subtracting the star and taking into account the \modif{difference of spectral index} differences is mandatory to reconstruct a reliable image of the environment.

Additionally, we verified that the conclusions remain unchanged when using other image-reconstruction algorithms than \texttt{MiRA}. We performed the same exercise using \texttt{Macim} and \texttt{Squeeze} with a regularization based on the Laplacian on the image. These algorithms rely on MCMC methods that were modified to handle the \texttt{SPARCO} approach. We obtained the same $\chi^2 (1.2)$ and the same conclusions (see Fig.~\ref{fig:img} bottom-right panel)

\subsection{Determination of chromatic properties}
\label{sec:pfit}

\begin{figure*}[!t]
 \centering
  \includegraphics[width=\hsize]{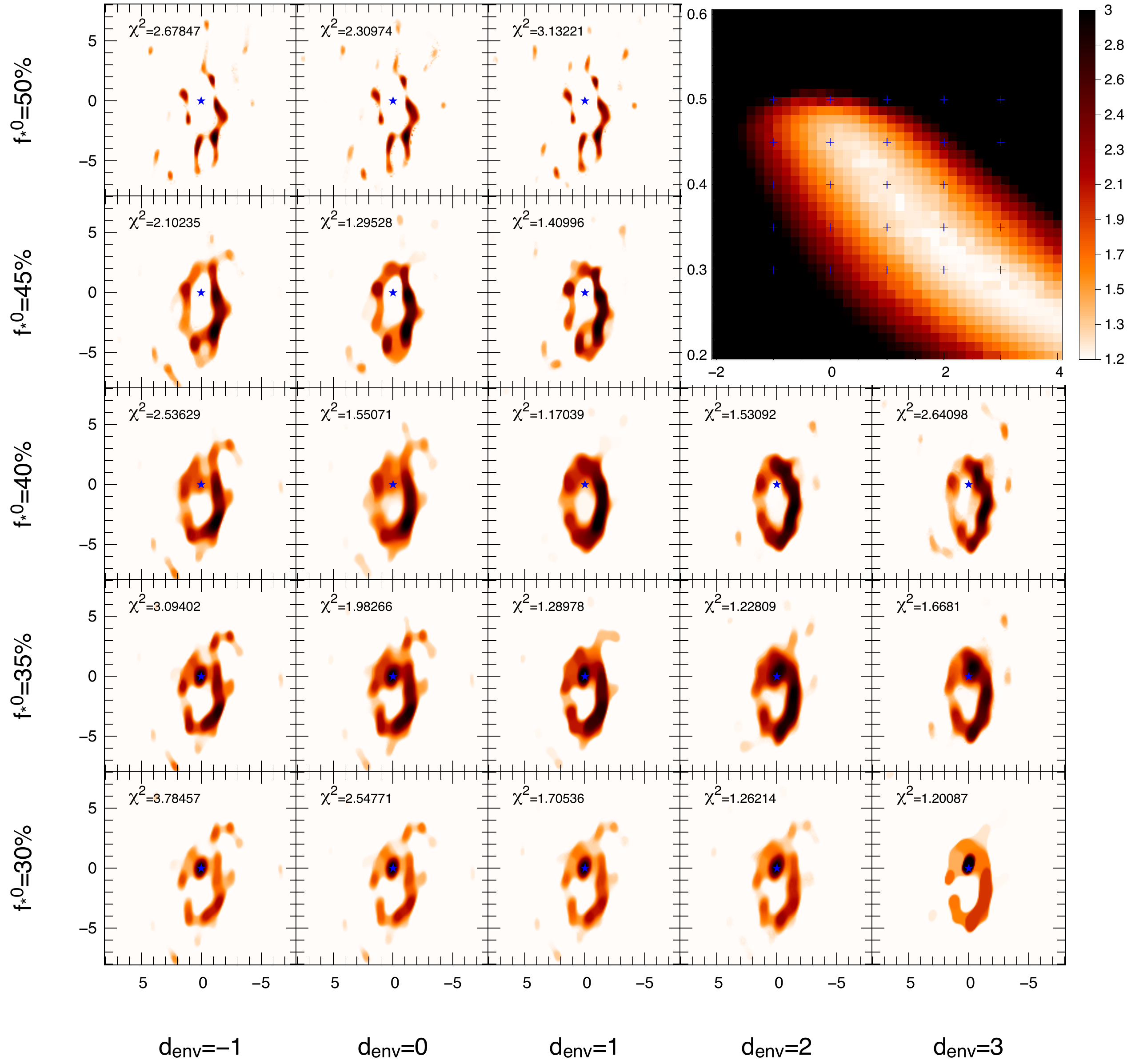}
 \caption{ \texttt{MiRA} image reconstructions as a function of the assumed chromatic parameters: the stellar-to-total-flux ratio $\fso$ and the dust spectral index $\denv$. The star is represented in red at the center of each image. The true values are $\fso = 40\%$ and $\denv = 1$. Top right: $\chi^2$ map of the reconstructions as a function of the chromatic parameters for reconstructions with the total variation regularization. All $\chi^2 \sup 3$ are represented in black. The blue crosses represent the location of the images in the $\chi^2$ map. We can clearly see that this map is degenerated. }
 \label{fig:mos}
\end{figure*}

The previous section presented a validation of the method in the ideal case where the chromatic parameters ($\fso$ and $\denv$) are known. In this section, we explore the effect of varying $\fso$ and $\denv$ (in other words:  the values used for the image reconstruction do not correspond to those used to simulate the dataset). This allows us to conclude whether it is possible to recover these parameters from the dataset.

The range of values for the environment spectral index $\denv$ was chosen to cover the temperatures of dust sublimation (from 2200\~K to 1100\~K, that is, $\denv = -1$ to 3 respectively). The range of values of the flux ratio $\fso$ was $0.2$ to $0.6$. The results are presented Fig.~\ref{fig:mos}. We found that the image morphology weakly depends on $\denv$ in the considered range of values. On the other hand, $\fso$ significantly influences the morphology of the reconstructed image.
If the stellar flux ratio is too high, the algorithm will make an image with a \modif{large} inner hole that affects the ring. In the \modif{opposite} case, if the star-to-total flux ratio ($\fso$) is too low, the algorithm will add flux at the star position. But the added flux is at the dust temperature. To compensate for this, the algorithm has to create artifacts in the image of the environment.

The top-right inset of Fig.~\ref{fig:mos} displays the $\chi^2$ value obtained at the end of image-reconstruction processes for the grid of $\fso$ and $\denv$. The $\chi^2$ is an indicator of the distance of the model (parameterized by the image pixel values, $\fso$ and $\denv$) to the dataset. The map shows a single global minimum. This validates the capability of reconstructing the image of the environment while simultaneously fitting the chromatic parameters ($\fso$ and $\denv$). Appendix~\ref{app:fitconv} demonstrates that this joint minimization is robust: it does not depend on the starting point or on the choice of the regularization.

However, the $\chi^2$-map \modif{shows} a correlation between the parameters $\fso$ and $\denv$. We verified that this degeneracy is not caused by regularization or the S/R ratio but is intrinsically linked with the interferometric data. 
\modif{As pointed out in Sect.~\ref{sec:gray}, artifacts in the reconstructed image compensate for an error on the chromatic parameters. This degeneracy seems to be carried out by these artifacts.}
\modif{Solutions to that degeneracy problem are} discussed in the next sections.

\section{Discussion}
\label{sec:dis}

\subsection{Need for spectrophotometry.}

We showed in Sect.~\ref{sec:pfit} that one can fit $\fso$ and $\denv$ when reconstructing the image, but that these parameters are strongly correlated.
Nevertheless, one can at least derive an upper limit to the star flux ratio ($\fso$) and a lower limit to the environment spectral index ($\denv$).

\modif{Improved image reconstruction is possible using additional constraints on $\fso$ and/or $\denv$ from other observations, for instance, spectrophotometry.} This is possible when the flux emitted by the central object can be safely extrapolated in the NIR, for instance when the central object is a star whose spectral type and extinction are known. In this case, spectrophotometric observations in the NIR allow one to recover $\fso$ and/or $\denv$. However, this simplistic interpretation of the SED should be applied with caution, as illustrated in Sect.~\ref{sec:HR5999}. 

\subsection{Effect of the gradient temperature in YSO's disks.}

Disks around young stellar objects display a temperature gradient in the radial direction. It has the same signature in the visibility as the star/environment chromatism \citep[e.g.,][]{2012ApJ...744...19K}. So far, this effect is not taken into account in SPARCO: we assume that the spectral index of the environment is not changing with radius. 

\begin{figure}[!t]
 \centering
      	\includegraphics[height=4cm]{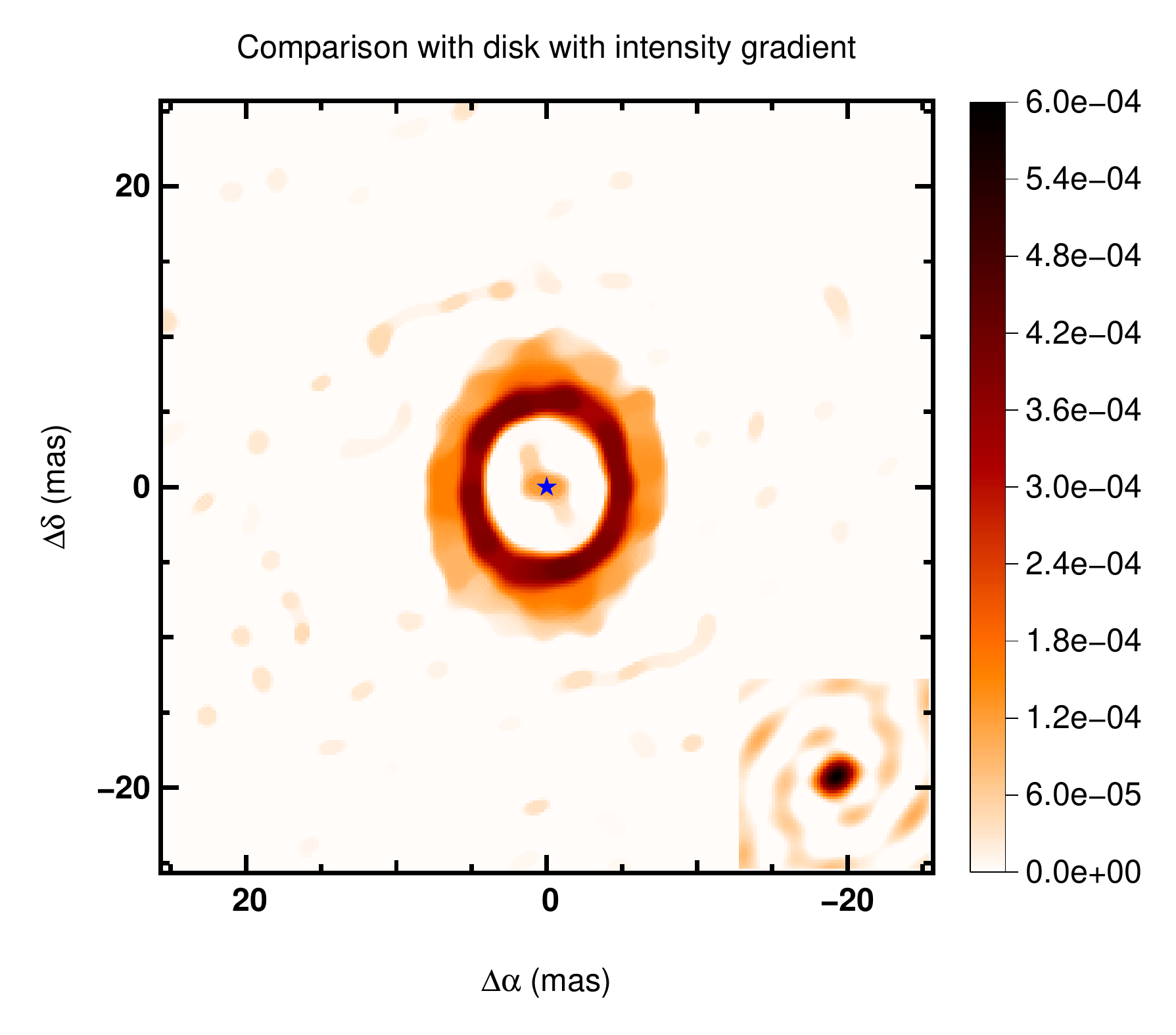}~
	\includegraphics[height=4cm]{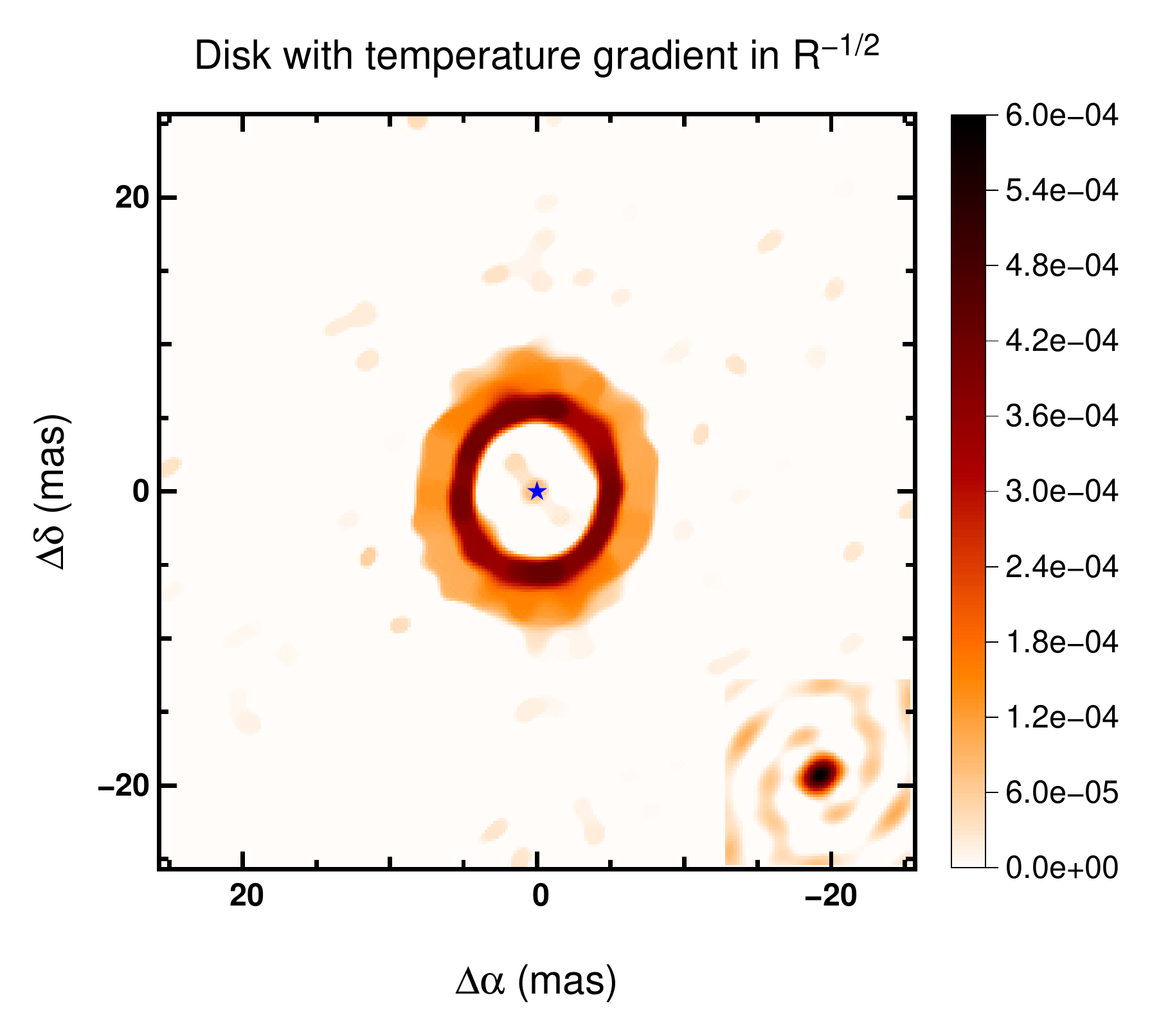}
              	\includegraphics[height=4.5cm]{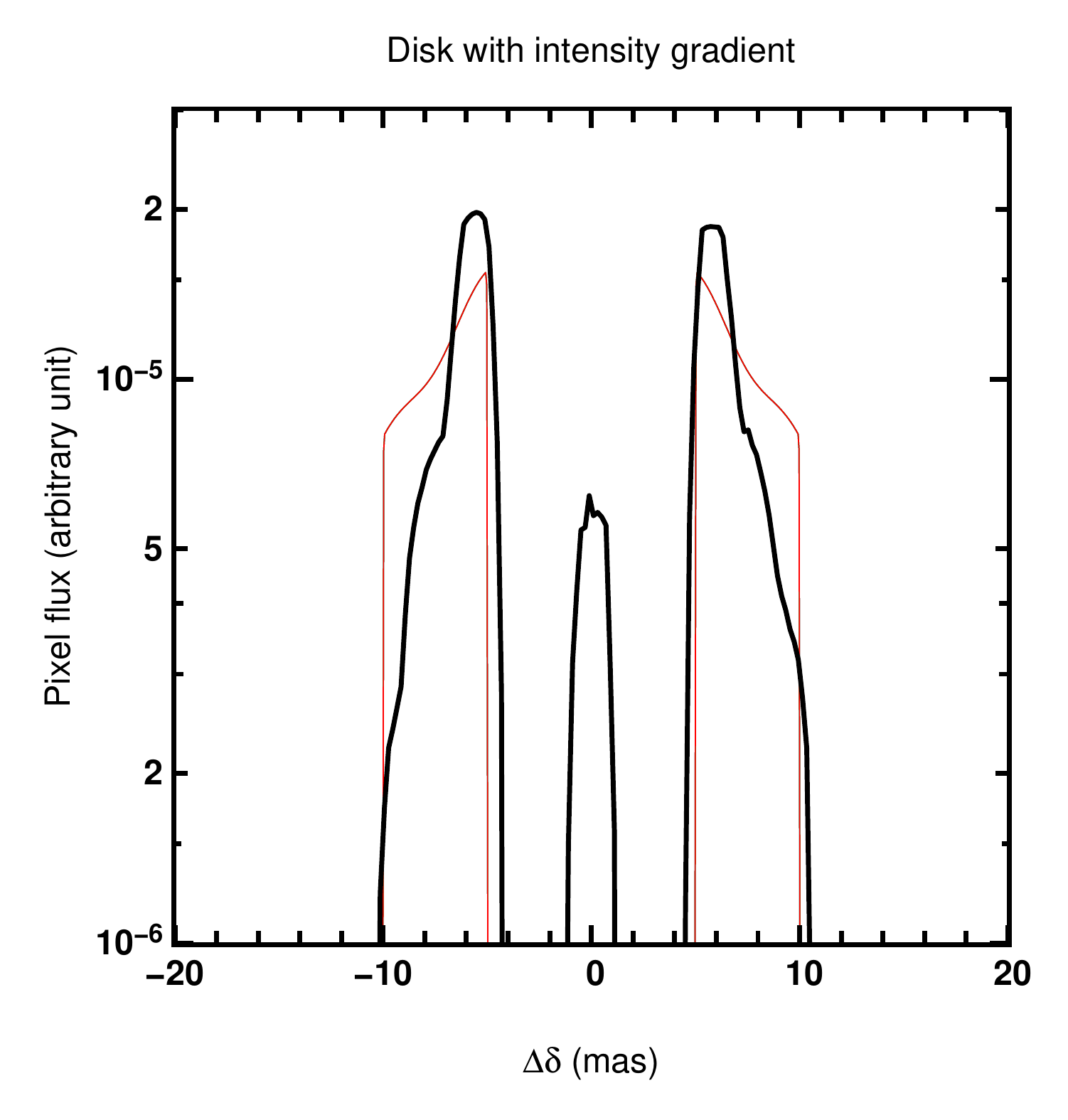}~~~
                \includegraphics[height=4.5cm]{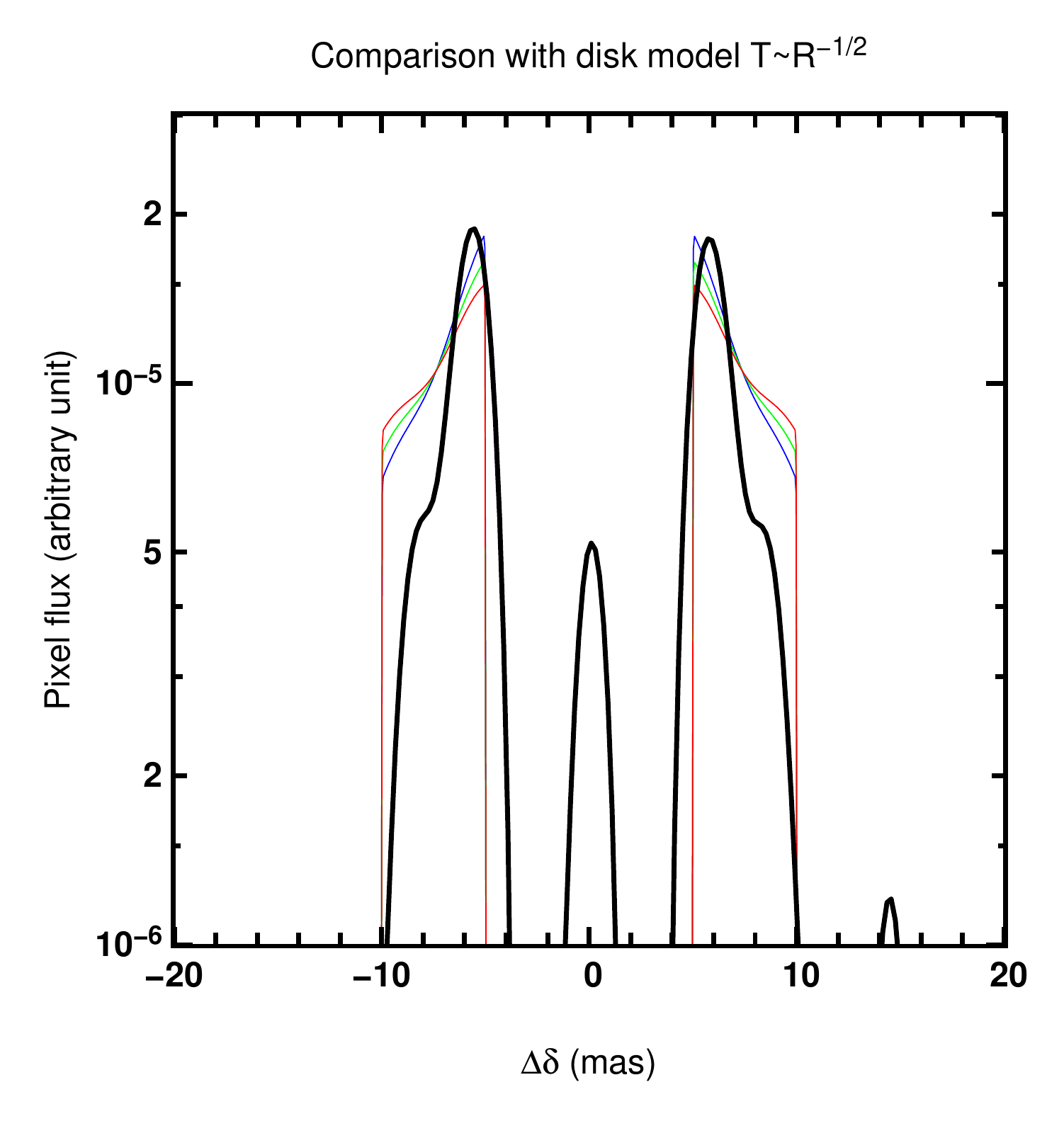}
 \caption{Reconstructions with an intensity gradient in the disk (left) and a temperature gradient in the disk $\propto R^{-\frac{1}{2}}$ (right). The reconstructions have a $\chi^2\approx1$. Bottom: cut of the reconstructed images (black) and of the disk models. The colors are for the different wavelengths in the right panel (blue: $1.55\mu$m, red: $1.8\mu$m). The flux level is given on a logarithmic scale.}
 \label{fig:acc}
\end{figure}

To test this, we built a disk model with a temperature gradient corresponding to a flared accretion disk with T $\propto \mathrm{R}^{-\frac{1}{2}}$\citep{1987ApJ...323..714K}. We also built a reference model with an intensity gradient matching the average intensity distribution of the first model. 
 The flux distribution is 50\% for the star and 50\% for the disk in the middle of the $H$ band. The disks extend from 5mas to 10mas. 
The temperature of the inner rim is 1500\~K.

Fig.~\ref{fig:acc} shows the reconstructed images with the SPARCO approach setting $\fso=0.5$ and letting $\denv$ free to vary. They have acceptable $\chi^2$ values ($\approx1.0$). The recovered $\denv$ corresponds to 1400\~K for the temperature gradient case and 1500\~K for the intensity gradient case. Interestingly, 1400\~K corresponds to the average temperature of the accretion disk. This is therefore a satisfactory value. The overall shape (size and radial extension) of the disk is qualitatively recovered in both cases. The inner part of the disk dominates the emission. The differences between the two models are significantly smaller than the dynamic \modif{range} of the image reconstructions. 

We also made the test with a flat accretion disk (T$\propto \mathrm{R}^{-\frac{3}{4}}$) and arrived at the same conclusions.
We conclude that the effect of a temperature gradient is negligible for image reconstruction of YSOs over the $H$ or the $K$band.

However, for multiband datasets we recommend \modif{splitting} the data into several spectral bands, which better adapts the flux description. In the mid- or far-infrared, where the star is negligible, we would need to modify the method for image reconstruction.

\subsection{Evolution of the parametric model.}
\label{sec:modif}
The parametric model of the star as an unresolved object is sufficient to detect interesting features in an object, but it may be inefficient for imaging objects that are more complex for observations with an interferometer with a sufficient resolution to resolve the central star. We can modify the algorithm to subtract other shapes than just an unresolved star. In Eq.~(\ref{eqn:gen}) we can replace $\widetilde{V}_*$ by any model with an analytical formula or even an image. For a binary we can upgrade this method to subtract both of the components and even find their positions.

In addition, the adaptation to other spectral bands for YSOs or to instruments with better spectral resolutions will pass through the modification of the flux laws for the star and the environment. We chose power laws because in the $H$ and $K$ bands (which are quite narrow) they reproduce the spectral behaviors of the two components well. Another parametric flux description can replace the current one in the Eq.~(\ref{eqn:chrom1}).



\section{Application on actual datasets: HR5999}

\label{sec:HR5999}

In this section we apply the \texttt{SPARCO method} to the actual dataset from \citet{HR5999} on HR5999, a Herbig Ae star \citep{1989A&AS...78....1T}. This dataset was obtained \modif{by} the VLTI/AMBER instrument, which is a three-beam combiner in the NIR. The object shows an excess in near- and mid-infrared spectral energy distribution. It was first imaged by \citet{HR5999} (see Fig.~\ref{fig:HR5999_imgrec}).
The photosphere contributes 22\% \modif{of the flux} in $K$ band. Moreover, there is an inner disk \modif{that} is marginally resolved (between $0.43\,\mathrm{mas}$ and $2.8\,\mathrm{mas}$) \modif{and} contributes 38\% of the flux. It might be interpreted as a gaseous disk or as refractory dust grains. The remaining flux comes from the inner dusty rim, which is believed to be at the dust-sublimation radius.

\begin{figure}[!t]
 \centering
      \includegraphics[width=5cm]{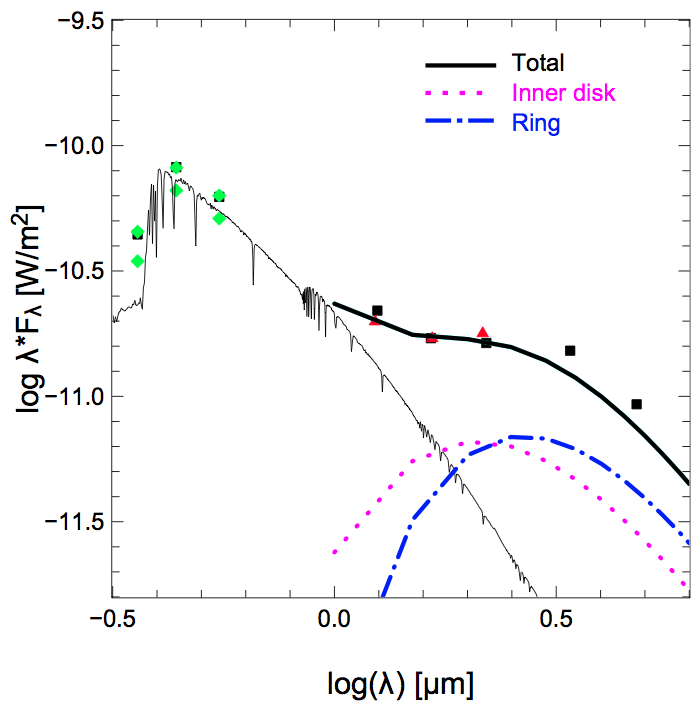}
 \caption{SED of HR5999 from \citet{HR5999}. We can see that in the $K$-band (at log(2.2$\mu$m) = 0.35), the emission is dominated by the environment which has two components. The star is weaker but has to be taken into account for the image reconstruction.}
 \label{fig:HR5999_SED}
\end{figure}

This object is quite complex because there is a third component that contributes to the NIR flux: the inner disk. Our method is not yet designed to answer this complexity. Nevertheless, we adopted a strategy in three steps to apply our method to this complex case: 
\begin{itemize}
\item{Reconstruction by subtracting the photosphere only from the image: using the SED (Fig.~\ref{fig:HR5999_SED}), we set the stellar to total flux ratio to the value of $\fso=22\%$ and the temperature of the environment to $1500\,\mathrm{K}$. The image (Fig.~\ref{fig:HR5999_imgrec} center) shows an unresolved component in the center and two patterns at both sides of the unresolved component. The subtracted flux is located at the position of the red star. As expected, the algorithm sees the unresolved flux, which is the inner disk, but has difficulties to reconstruct the environment correctly. We note that the position angle is similar to that found in \citet{HR5999}.}
\item{Reconstruction by subtracting the photosphere and the inner disk: the flux in the unresolved component is now set to $60\%$ (as show in the SED). The environment is still set to be at $1500\,\mathrm{K}$ \modif{($\denv \approx 1$)}. There is no feature at the center of the image (see Fig.~\ref{fig:HR5999_imgrec}). We can see the inner dusty disk rim. The image looks cleaned in the center because the fluxes \modif{(star + inner disk)} that we are subtracting are not necessarily unresolved \modif{(i.e., the inner disk was marginally resolved)}.}
\item{Fit of the chromatic parameters in the reconstruction process: if we did not have the information on the photometry on these objects, we would try the extended method described in Sect.~\ref{sec:pfit}. We converge to subtracted flux $\fso$ of 40$\%$ and a relative spectral index $\denv$ of -1 (which translates into a temperature of $2100\,\mathrm{K}$ if we assume that we only subtract the photosphere). We subtract the photosphere and half of the inner disk. The rest of the inner disk is considered as resolved.}
\end{itemize}

\begin{figure*}[!t]
 \centering
      \includegraphics[width=7cm]{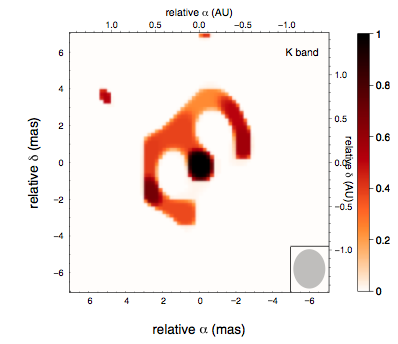}
  \includegraphics[width=5.5cm]{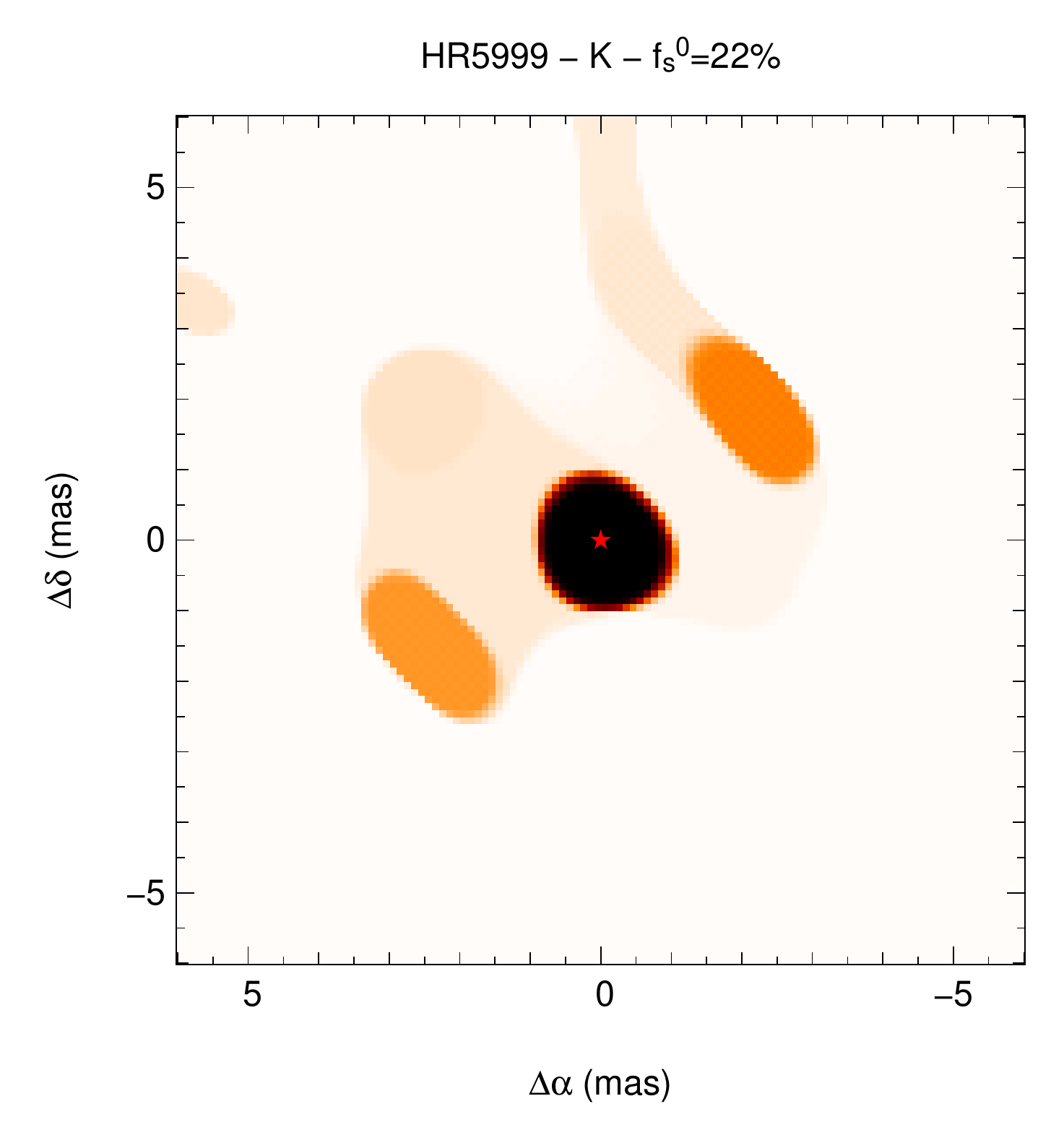}
                 \includegraphics[width=5.5cm]{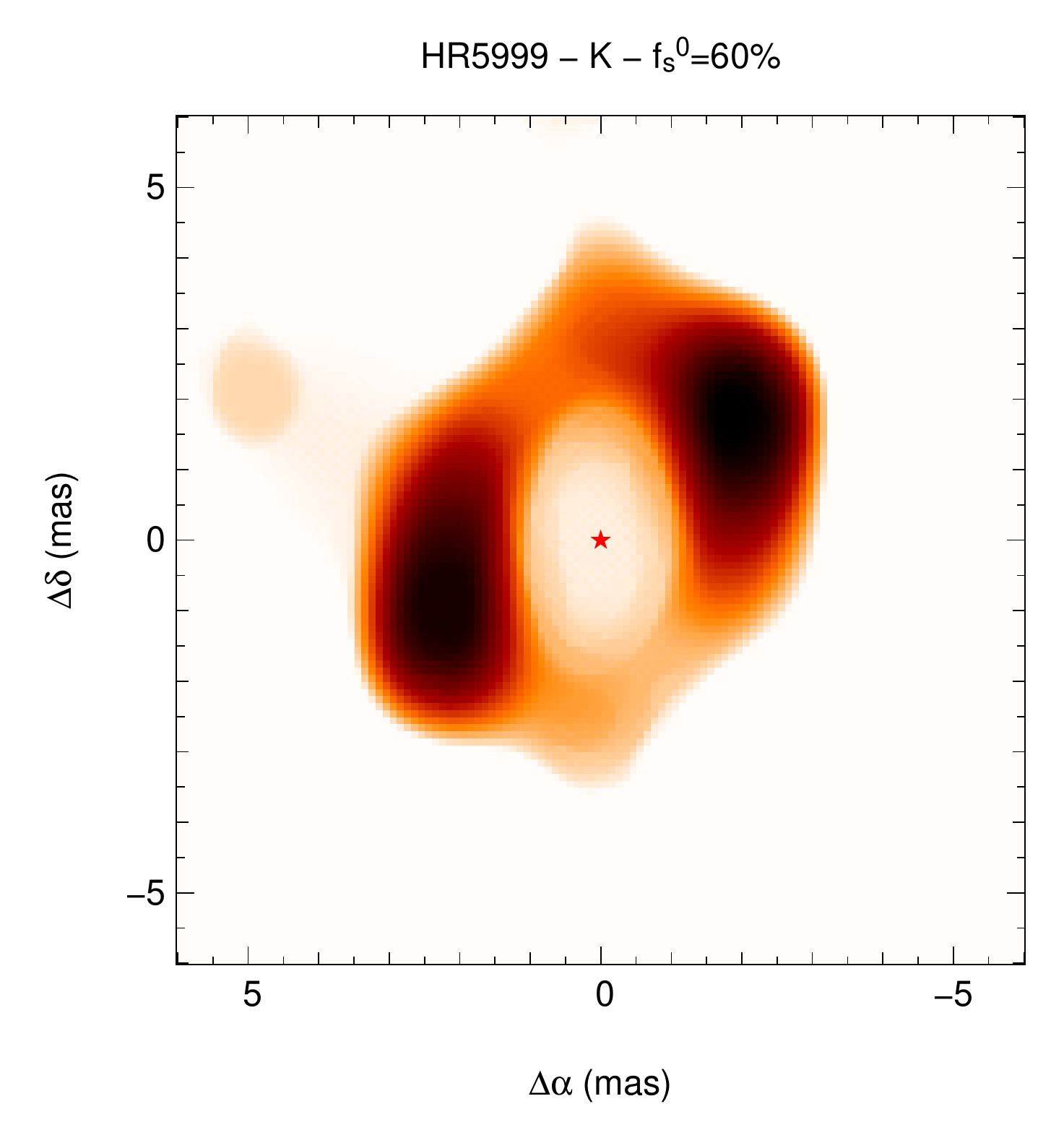}
 \caption{Image reconstructions on HR5999. Left: \citet{HR5999}. Center: \texttt{SPARCO} with 22\% of flux in the central red star. This corresponds to the amount of flux from the photosphere. Right: \texttt{SPARCO} with 60\% of flux (corresponds to the photosphere and inner disk contribution) in the central star.}
 \label{fig:HR5999_imgrec}
\end{figure*}

The first conclusion of the application of \texttt{SPARCO} on this dataset is that we retrieve the inner dust rim if we subtract the photosphere and the inner disk. We have to be careful to correct the spectral index of the environment $\denv$, which depends on the temperature of the unresolved flux. If we only subtract the photosphere, \texttt{SPARCO} reconstructs the unresolved flux from the inner gaseous disk. 

Second, if we let the method fit the chromatic parameters in the reconstruction process, it finds an $\fso$ that is larger than the photosphere contribution and a spectral index $\denv$ that indicates a temperature of more than $2000\,\mathrm{K}$. This is higher than the dust sublimation temperature of $1500\,\mathrm{K}$. 
\modif{This higher temperature is explained by the fact that the unresolved component around the star has a spectral index in K band "cooler" than the star itself, so the difference in spectral indices favored from the chromatic effect in the interferometric data is lower. This will automatically increase the derived temperature of the extended component if we still assume that the unresolved component is in the Rayleigh-Jeans regime.}

However, \citet{HR5999} suggested that the origin of the emission might be refractory dust grains or a gaseous disk. This indicates that the hypothesis of the photosphere alone as the unresolved component \modif{can be excluded}. 

The method is sensitive to an unresolved flux that does not come from the photosphere alone. The SED clearly indicates an inner component, as suggested in \citet{HR5999}.

\section{Conclusion}
\label{sec:sum}

\modif{For image reconstruction based on interferometric data with spectral dispersion, visibilities are determined not only by the geometry of the object, but also by the (differential) spectral slope of its components. }
We developed a method that includes the knowledge that we have on \modif{the relative spectral behaviors of two components} to reconstruct the intensity distribution of the \modif{extended} one. This method allowed us to improve the $\chi^2$ by one order of magnitude in our validation.

The \texttt{SPARCO} method includes an analytical description of the stellar contribution and the chromatic ratio between the star and its environment in the image-reconstruction algorithms used in optical interferometry. The first component is modeled parametrically and an image is reconstructed for the second one.

\texttt{SPARCO} was used correctly on a young stellar object with the following hypothesis: 

\begin{itemize}
\item \textit{The environment spatial \modif{distribution} is assumed to be wavelength independent.} The chromatic dependence of visibilities only arises because the star has a different spectral dependence from the environment. This also
implies that the environment has \modif{the same spectral dependence in the whole image}. 
\item \textit{The star is very close to the parametric description given in the method.}
\item \textit{The fluxes ($F_{\lambda}$) of the components are approximated by power laws as a function of the wavelength. Since the chromatic parameter space can be degenerate when considering interferometric constraints alone, obtaining independent spectrophotometric observations helps to retrieve an image of the observed environment. Despite the low sensitivity of the method to the choice of regularization, one has to be careful when choosing the value of the hyper-parameter $\mu$.}
\end{itemize}

The method is simple but can be adapted to more complex models. As demonstrated in Sect.~\ref{sec:HR5999}, if the object is more complex some more information needs to be added to the model (add a parametric disk for the inner part of the object). 
In Sect.~\ref{sec:modif} we showed that we can easily model the parametric part by a uniform disk, a binary, and even an image.

This tool is complementary to other observations of the target, especially with spectrophotometric observations. Since the chromatic parameter space can be degenerate, these types of observations are very important to correctly retrieve an image of the observed target. It is difficult to retrieve them by simultaneous fitting.

The \texttt{SPARCO} method will be intensively used on a Large Program dataset on Herbig Ae/Be stars gathered by PIONIER at the VLTI.
Unraveling the image of the close environment of young stars will help us to constrain the effects of the inclination of the inner regions of YSO and therefore to detect early signs of planet formation very close to the star. This method cannot be applied only to young stars, but can also be used in any system whith a point-like source and that displays a different spectral behavior from the rest of the emitting material in the optical such as active galactic nuclei.


\modif{Algorithms that implement a fully polychromatic approach are currently being
developped, e.g. \texttt{MiRA} \citep{Mira3D} and \texttt{Squeeze} (Baron et al. in prep.). Combining the \texttt{SPARCO} approach with polychromatic reconstruction
will allow imaging any stellar environment with limited perturbation from
its star, and greatly enhance our capability to study variations in the
environment morphology (e.g. temperature gradient in the disk).}



 \begin{acknowledgements}
This work is supported by the French ANR POLCA project (Processing of pOLychromatic interferometriC data for Astrophysics, ANR-10-BLAN-0511). 
\\
We acknowledge Myriam Benisty for fruitful discussions and for providing us the dataset on HR5999.
\\
We acknowledge Christophe Pinte, Michel Tallon and Isabelle Tallon-Bosc for interesting discussions that led to this paper.
\end{acknowledgements}

\bibliographystyle{aa}
\bibliography{biblio}

\appendix

\section{Reconstructions with different signal-to-noise levels.}

\begin{figure}[!t]
   \centering
   \includegraphics[width=9cm]{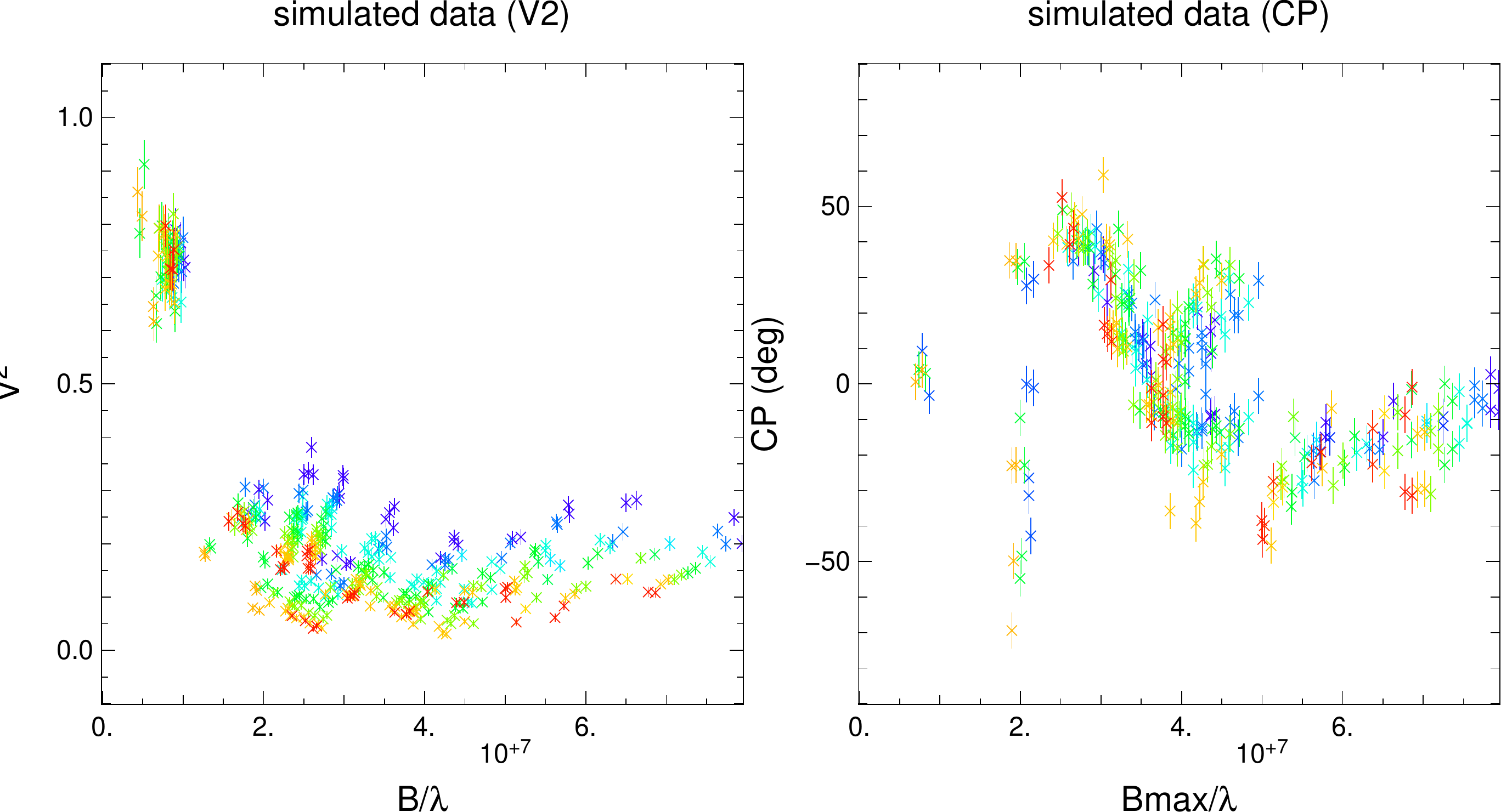} 
   \includegraphics[width=9cm]{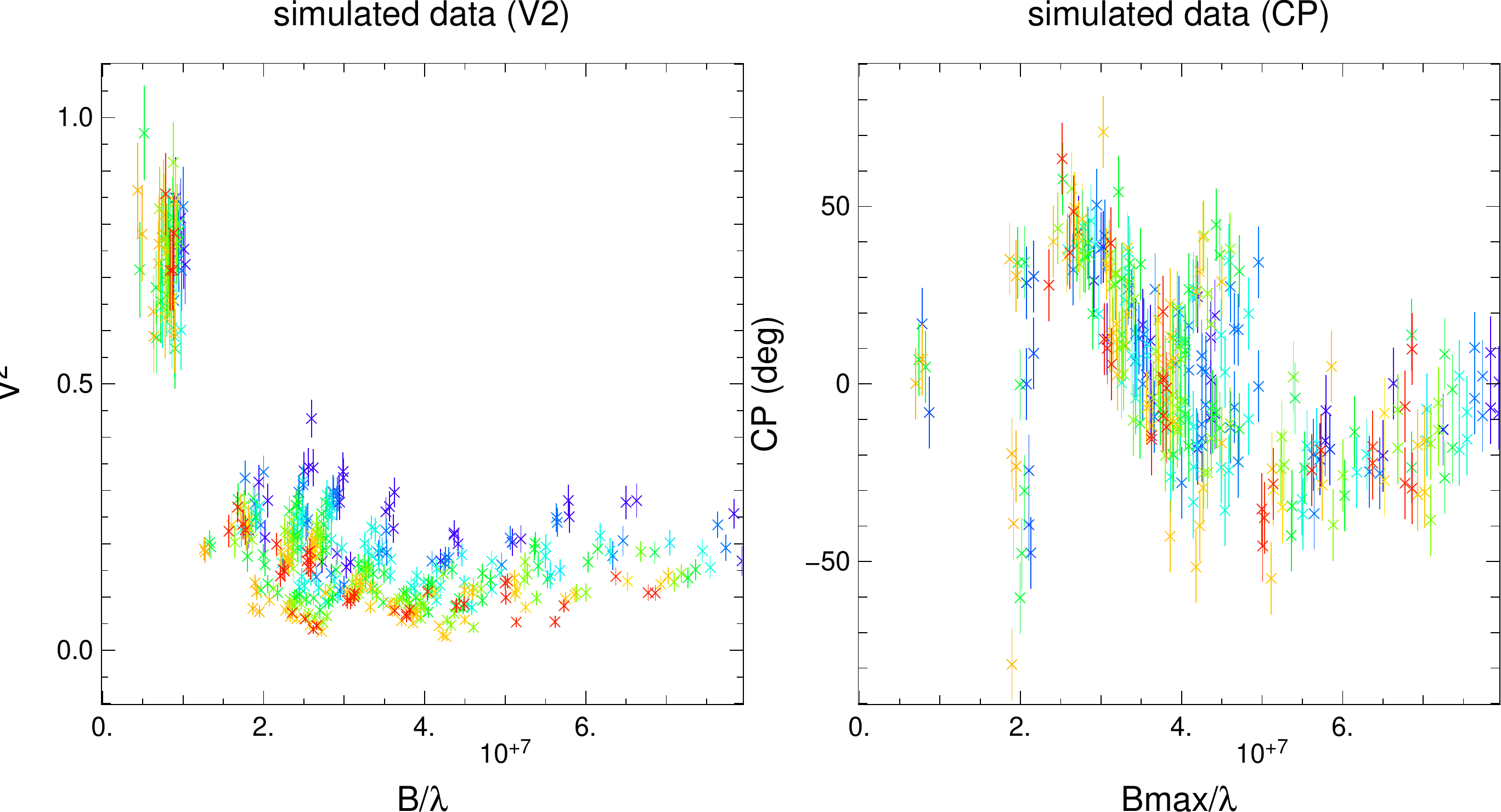}
      \caption{Intermediate and high photon noise regimes, left: squared visibilities, right: closure phases.}
      \label{fig:noisy}
   \end{figure}

We show in Fig.~\ref{fig:noisy} the simulated datasets with intermediate and high SNR. 

\begin{figure}[!t]
\centering
\includegraphics[height=3.8cm]{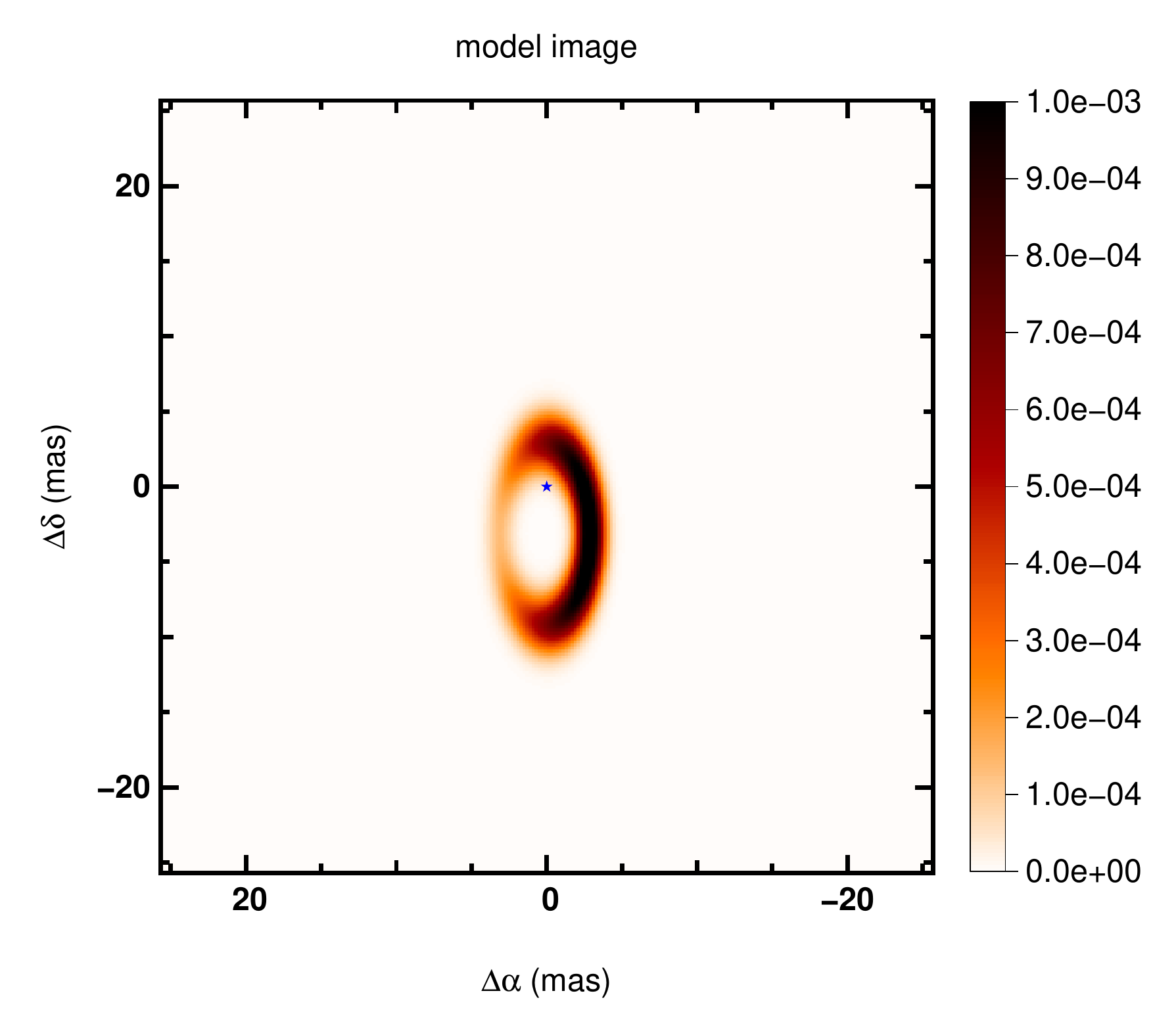}
\includegraphics[height=3.8cm]{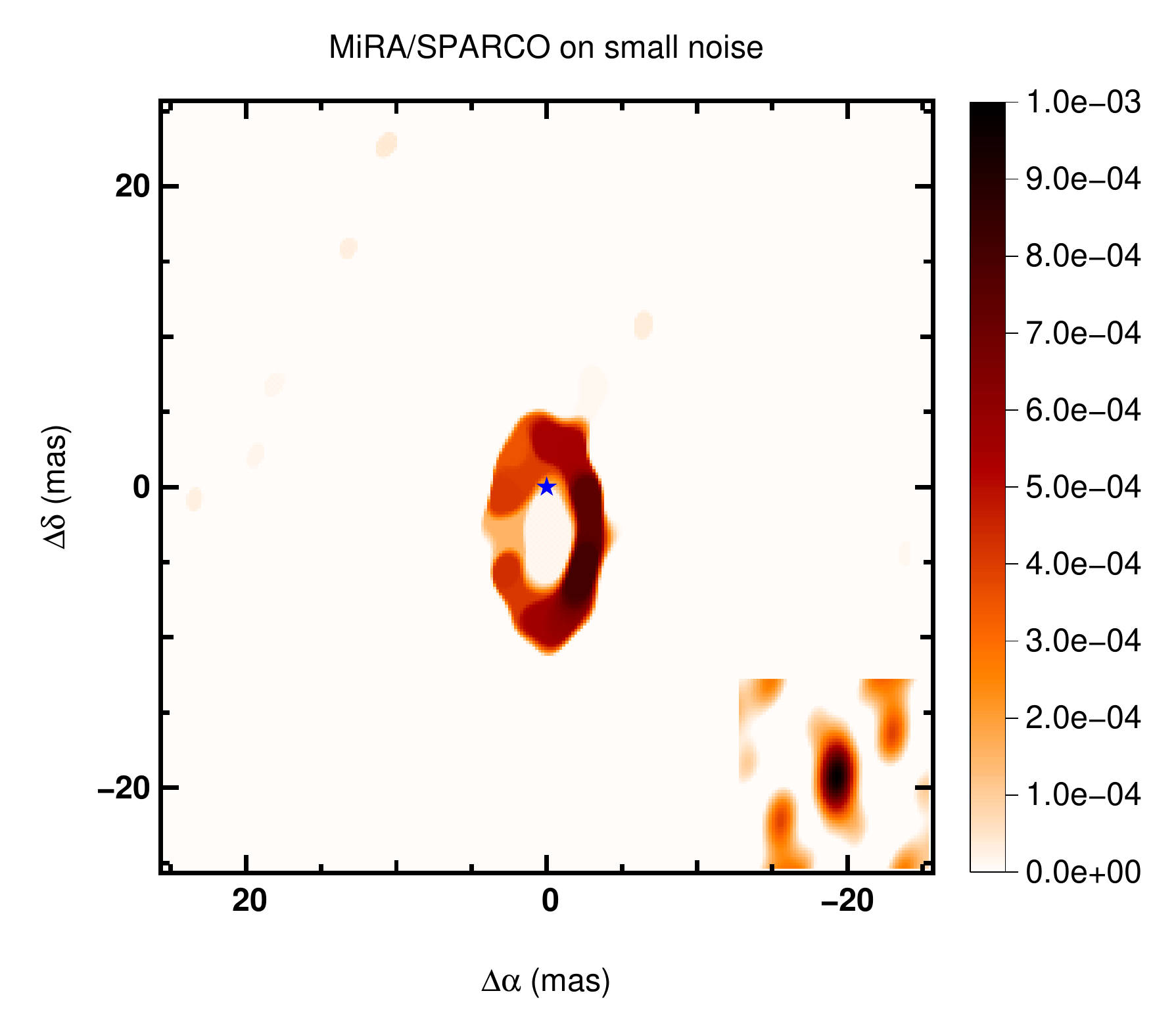}
\includegraphics[height=3.8cm]{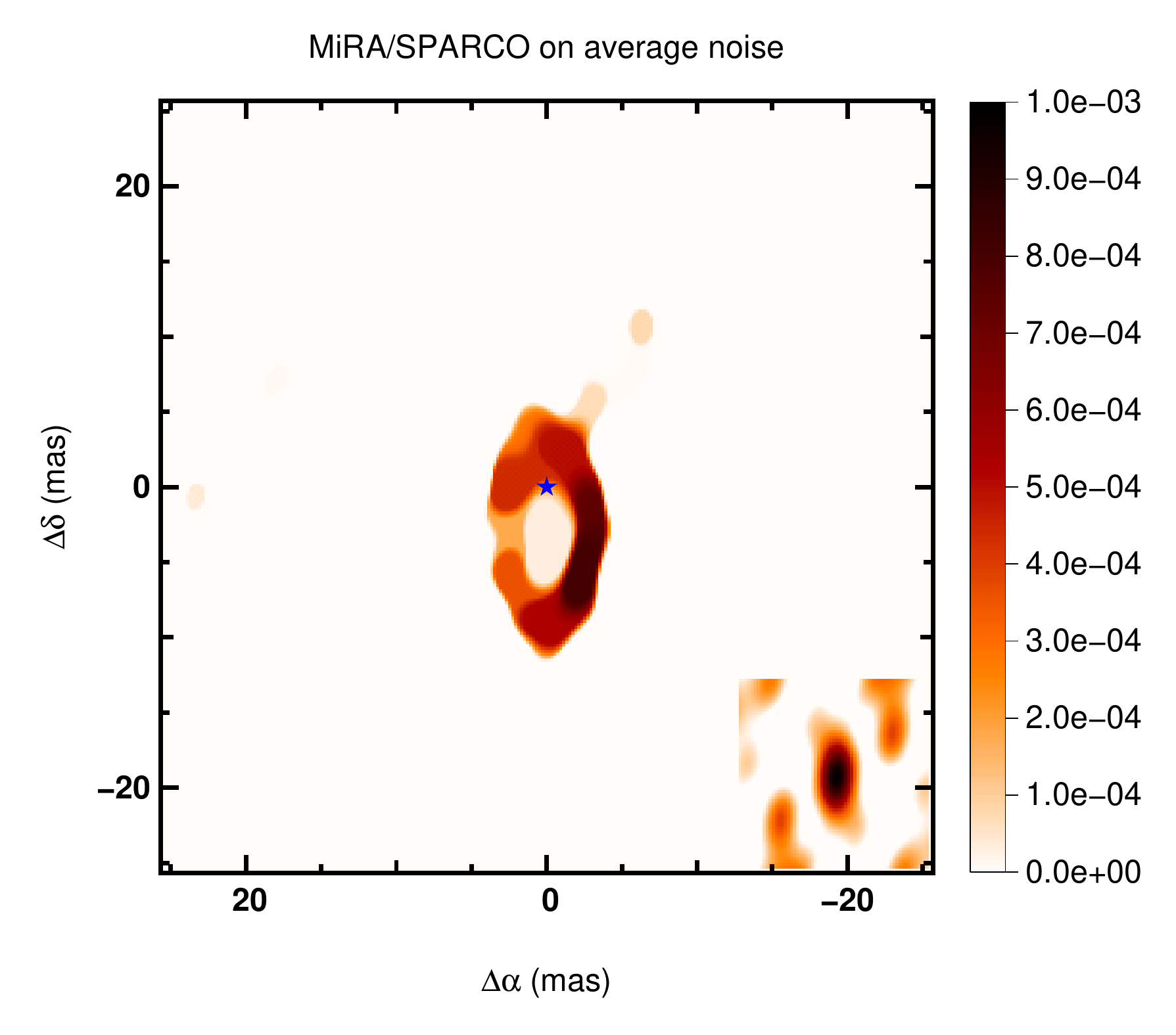}
\includegraphics[height=3.8cm]{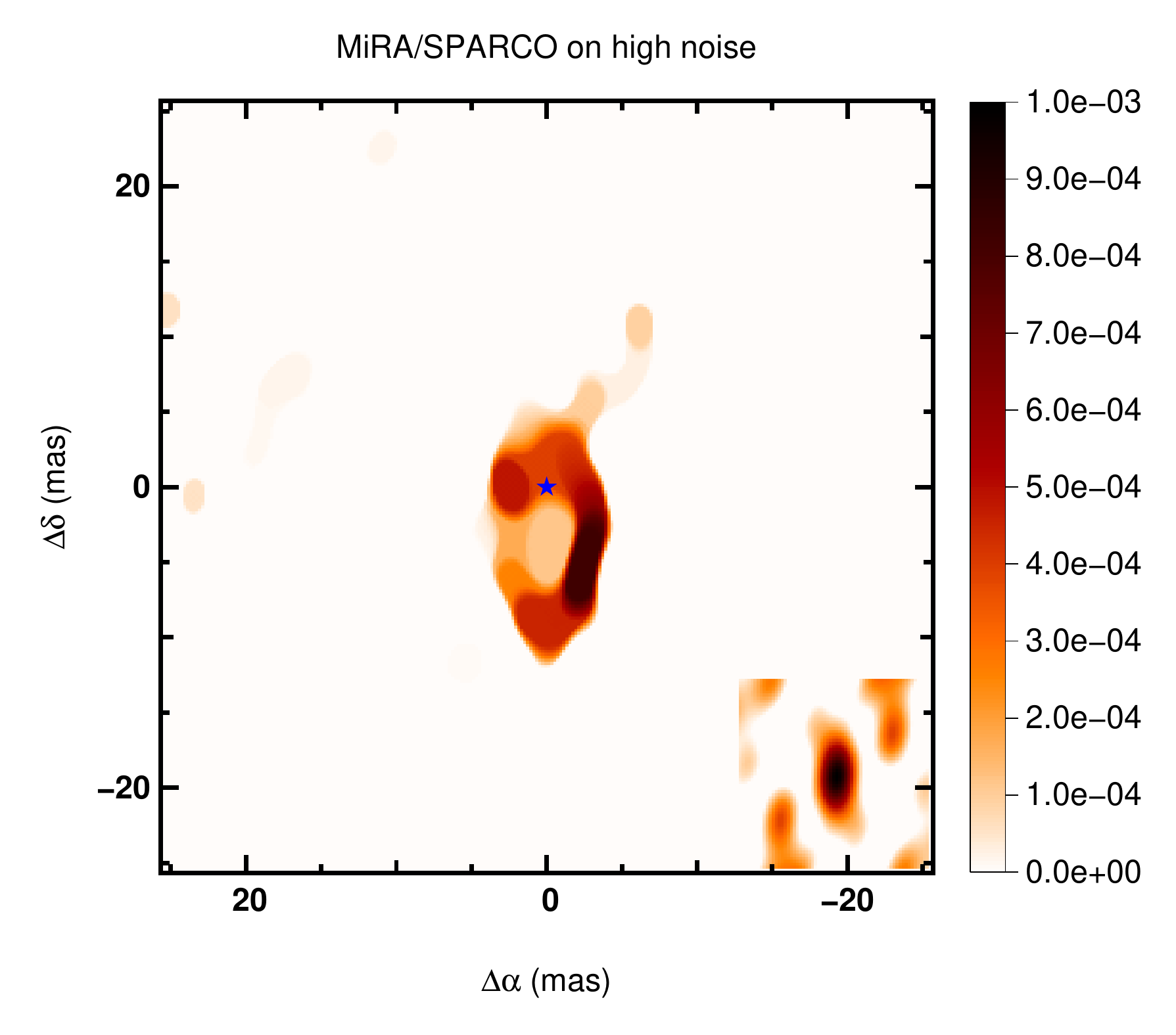}
\caption{\texttt{MiRA} image reconstructions with \texttt{SPARCO}. The top-left image is the original model image. In the top right we show the reconstruction with the low-photon noise level. In the bottom left corner there is an image reconstruction with the average photon noise level. In the bottom-right corner the reconstruction was made with the high photon noise level. The red star represents the central star in all the images. With this method, the star is not represented in the image as flux in the pixels. In the bottom-right corner of each image the dirty-beam is represented. The dirty-beam is the equivalent of the point spread function in interferometry. It is computed from the \uv-plan.}
\label{fig:imgrec}
\end{figure}

We compare the reconstructions with different SNR ratios. The starting point of the reconstructions is an image with flux in one central pixel alone. The results with the three different noise levels are presented Fig.~\ref{fig:imgrec}. We used the chromatic parameters corresponding to those we set in our model.

The ring is well reproduced in the sense that its size, width, and orientation are good. Furthermore, as expected, the algorithm is able to shift the environment with respect to the star.

Moreover, the azimuthal modulations are well reproduced even if there is a quality variation with the noise. It does not interpret the closure phases by asymmetric shapes. For the noise contribution, the global shape of the object and the details are well represented even with the high photon noise level. We conclude that there is no strong limitation on retrieving the global shapes and asymmetries of the objects in the PIONIER operating regime (see Fig.~\ref{fig:imgrec}).

\section{Choice of the regularization}
\label{app:rgl}

\subsection{Choice of the hyper-parameter $\mu$}

\begin{figure*}[!t]
\centering
\includegraphics[width=5.5cm]{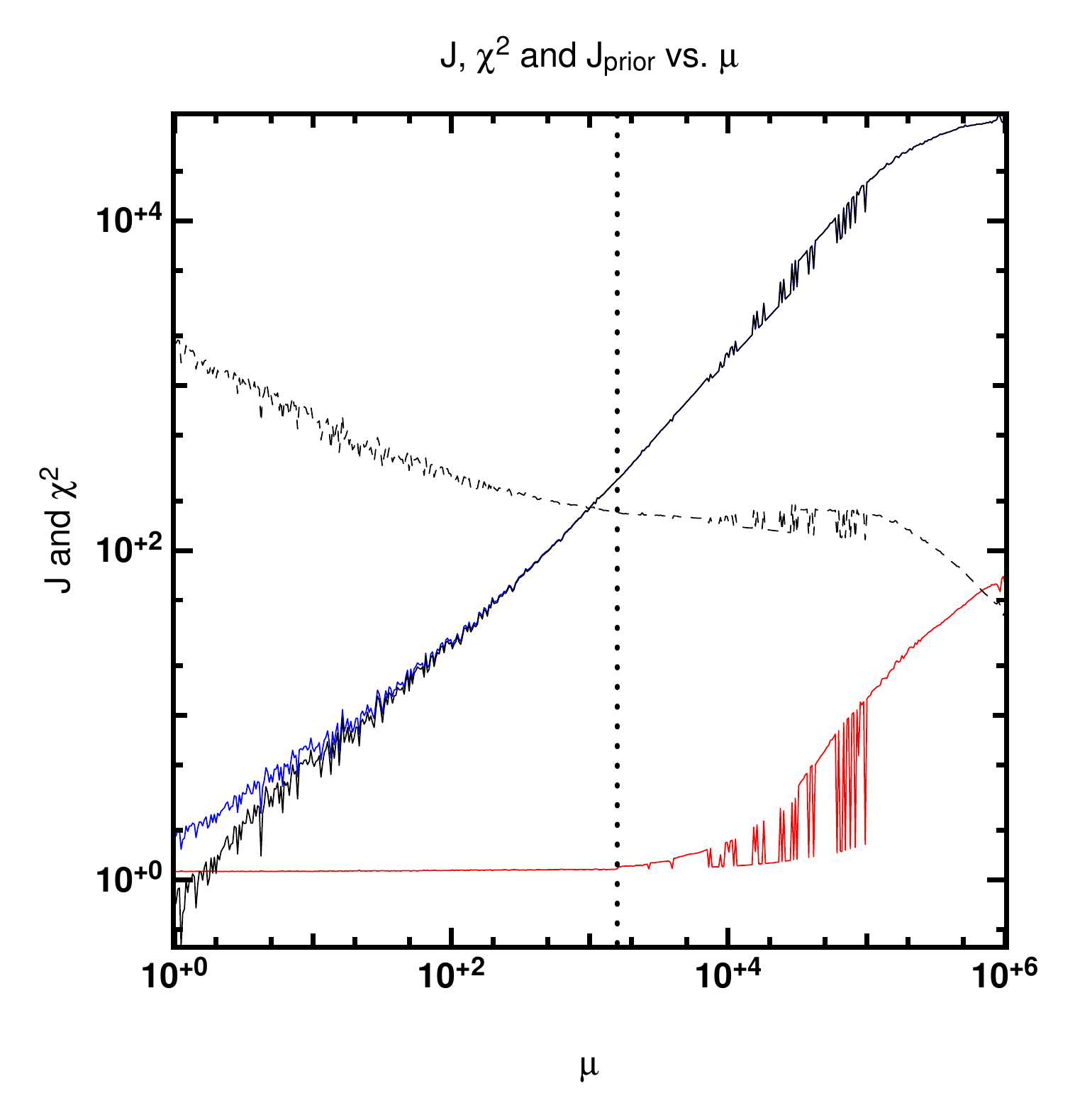}
\includegraphics[width=5.5cm]{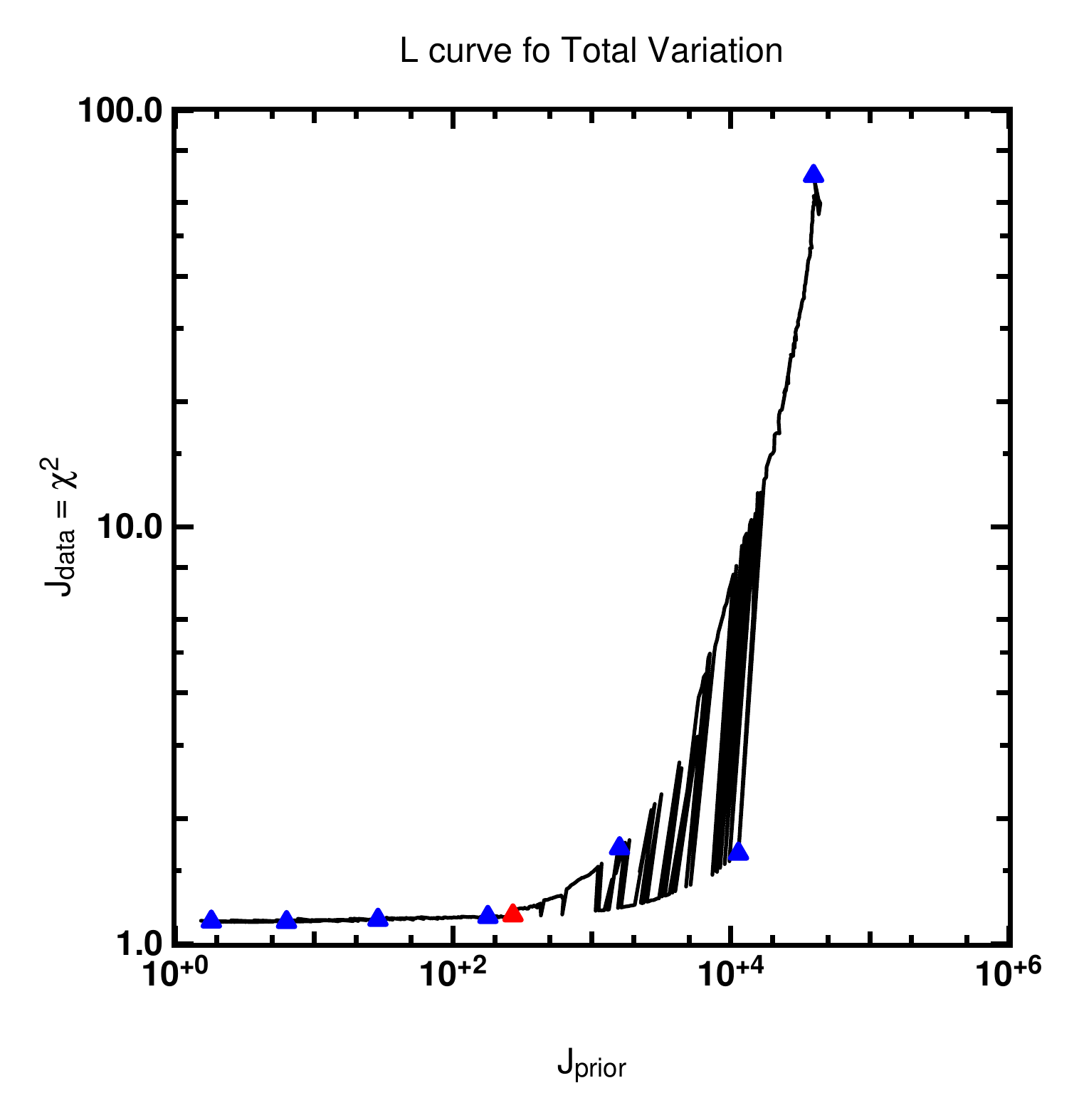}
\includegraphics[width=5.5cm]{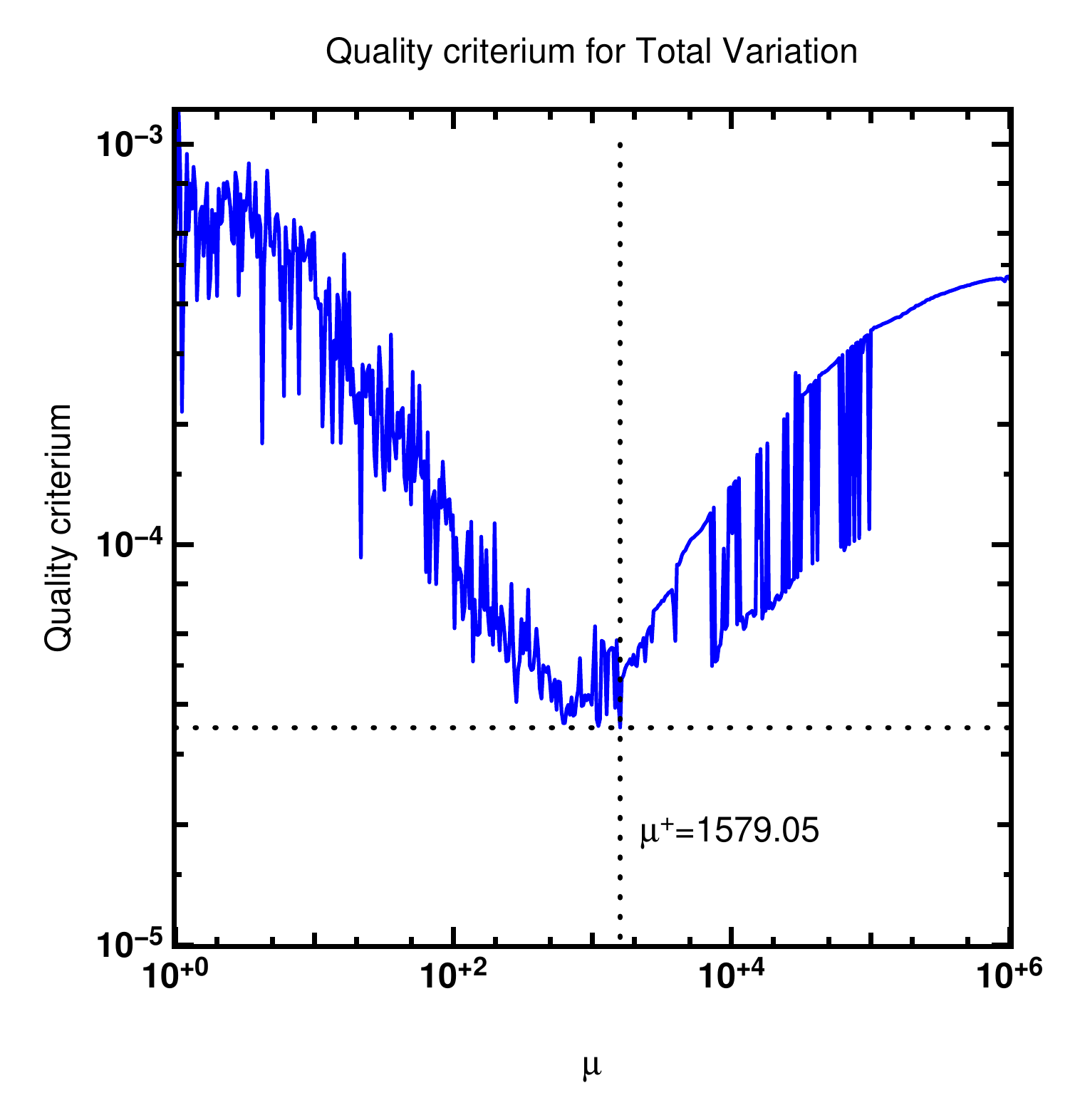}
\caption{Tuning the hyper-parameter $\mu$ for the total-variation regularization with \texttt{MiRA/SPARCO}. Left: cost functions vs. the hyper-parameter $\mu$. Red line: $\chi^2$ (=$\mathcal{J}_{\mathrm{data}}$). Black dashed line: $1000\times\mathcal{J}_{\mathrm{prior}}$ (multiplied by 1000 for better graph visibility). Black solid line: $\mu \mathcal{J}_{\mathrm{prior}}$. Blue line: $\mathcal{J} = \mu \mathcal{J}_{\mathrm{prior}} + \chi^2$. The vertical dotted line corresponds to the best regularization weight $\mu^+$ found by the quality criterium. Center: The L curve is one of the criteria to choose the regularization weight $\mu$. Black line: the values of $\mathcal{J}_{\mathrm{data}}$ vs. $\mathcal{J}_{\mathrm{prior}}$ for a range of $\mu$ values from $1$ to $10^6$. Blue triangles: points where $\mu$=\{$1, 10, 10^2, 10^3, 10^4, 10^5, 10^6$ \}. Red triangle: the value where $\mu = \mu^+$. Right : the quality criterium graph is the distance between the pixels of the reconstructed image to the real (here the model) one. Blue line: the value of the quality criterium vs. the hyper-parameter $\mu$. Dotted line: the position of the minimum of the curve and the corresponding value of $\mu = \mu^+$. }
\label{fig:rgl}
\end{figure*}

We reconstructed the images using the total variation regularization, because this regularization has been shown to be well-adapted to many types of objects \citep{Renard}. As mentioned in Sect.~\ref{sec:imgrec_base}, we need to tune the hyper-parameter $\mu$ in Eq.~(\ref{eqn:f}) to set the weight of the regularization. This needs to be done carefully to obtain a strong regularization and to still stick to the data. We present in Fig.~\ref{fig:rgl} three graphs that helped us to find a good value of the hyper-parameter $\mu$. Ultimately, the best way to find the best value for $\mu$ (which is noted $\mu^+$) is to compare the reconstructed image with the real one (when the model is available). The images were normalized to a total flux of unity. We summed the differences of pixel values throughout the image. This is called the quality criterium (QC):
\begin{equation}
\mathrm{QC}= \|{x^{\mathrm{rec}}_{\mu} - x^{\mathrm{ref}}}\|^2,
\label{eqn:QC}
\end{equation}
where $x^{\mathrm{rec}}_{\mu}$ is the reconstructed image pixel with the regularization weight $\mu$, and$x^{\mathrm{ref}}$ are the pixels of the model image.
The disadvantage of this method is that we must have the ``true'' image, which is not possible with real data. That is why this QC method indicates whether the definition of $\mu^+$ found by another method is accurate. We made the QC analysis for our model with reconstruction using \texttt{SPARCO/MiRA} with total-variation regularization (see Fig.~\ref{fig:rgl} right). Each reconstruction starts with all the flux in one central pixel. We found $\mu^+\approx1500$. The right graph in Fig.~\ref{fig:rgl} shows a wide range of $\mu$ that corresponds to a good image reconstruction (typically $500 < \mu<2000$). 

For real datasets, we have no access to the real image. We aim for an image that is well regularized but still sticks to the data. The $\mathcal{J}_{\mathrm{data}} (=\chi^2)$ curve (Fig.~\ref{fig:rgl} on the left) increases sharply with $\mu>\mu^+$. 

This is confirmed by the L curve $\mu^+$ determination (see Fig.~\ref{fig:rgl} center). We plot $\mathcal{J}_{\mathrm{data}}$ vs. $\mathcal{J}_{\mathrm{prior}}$ in a log log graph. The $\mu^+$ location should be in the corner of the curve. It actually is just in front of a strong deviation from the minimum of $\mathcal{J}_{\mathrm{data}}$.

Two regimes are present for $\mu > \mu^+$. We can see them on all the graphs of Fig.~\ref{fig:rgl}. This is the effect of the gradient-descent method that remains in local minima for some $\mu$ values and moves to a better minima for others. To correct that we could start with a prior image that is closer to what we expect to see. 
This effect should not be noticeable with \texttt{Macim} and \texttt{Squeeze} since the Monte Carlo method is less likely to remain in local minima.

In all the following reconstructions we took $\mu$ to 1500. To allow comparisons we did not change its value between the different methods.

\subsection{Effect of regularization on chromatic-parameter $\chi^2$-map.}
\label{app:rgldis}
\begin{figure}[!t]
 \centering
      \includegraphics[height=4.1cm]{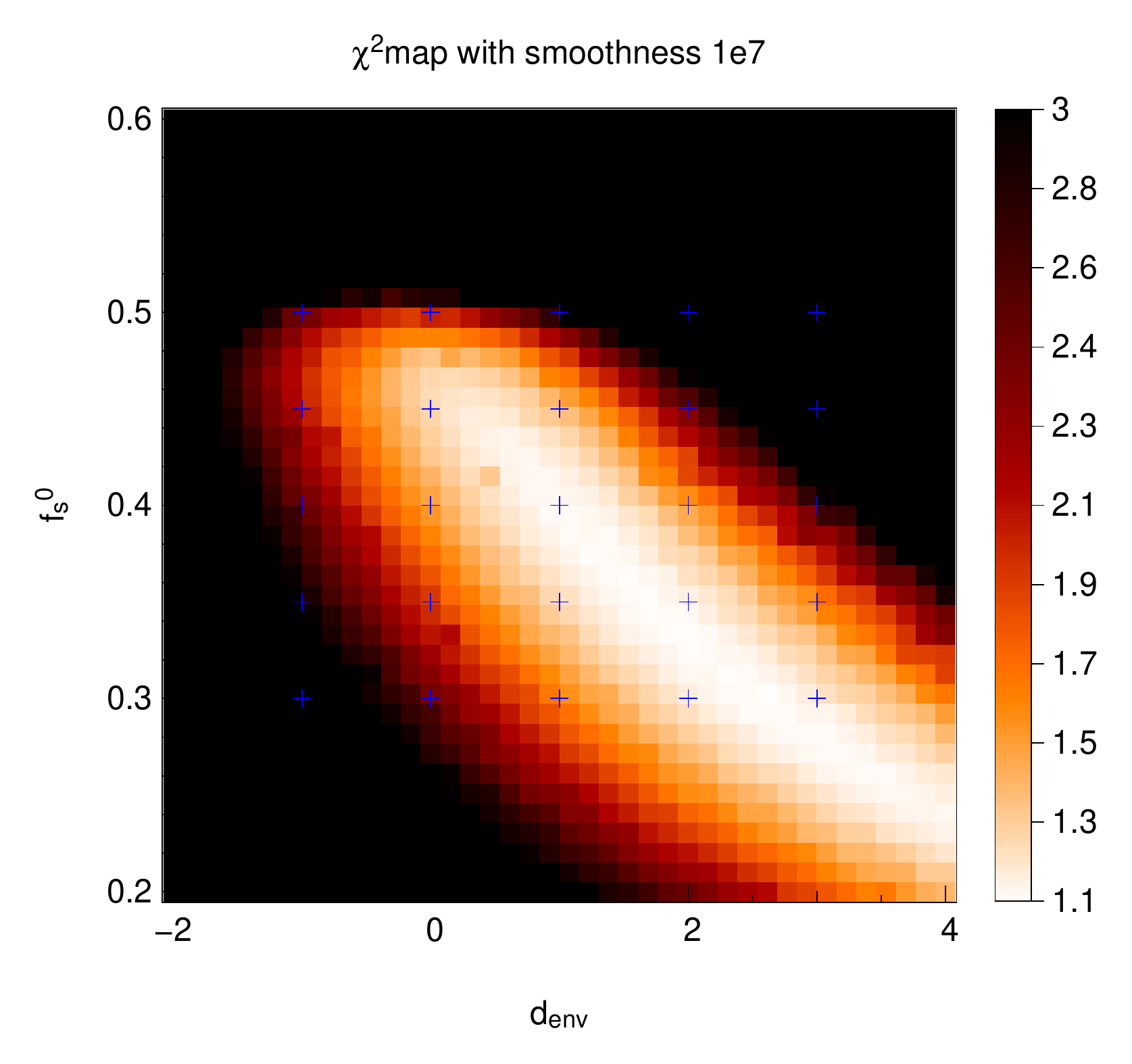}
      \includegraphics[height=4.1cm]{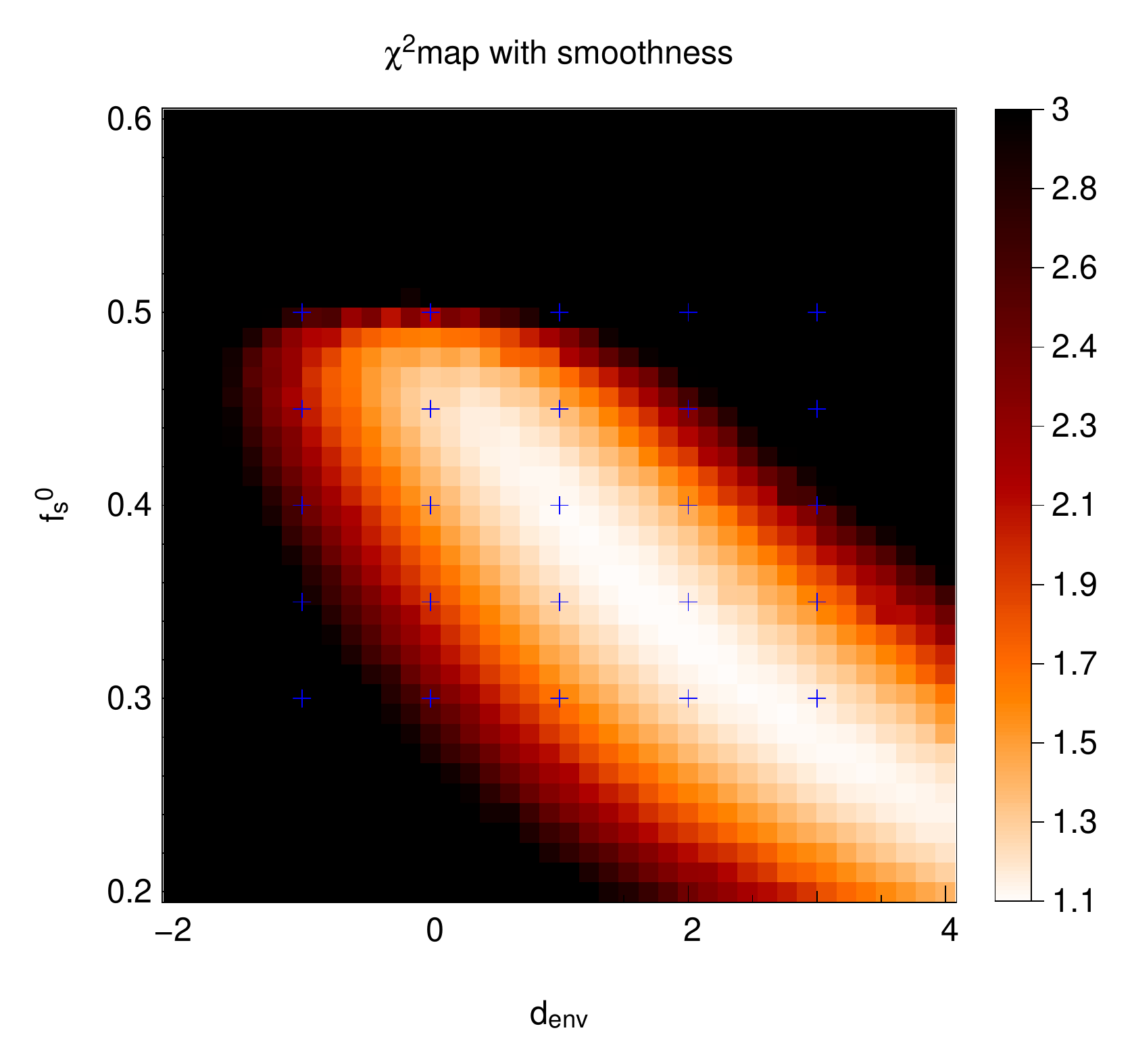}
      \includegraphics[height=4.1cm]{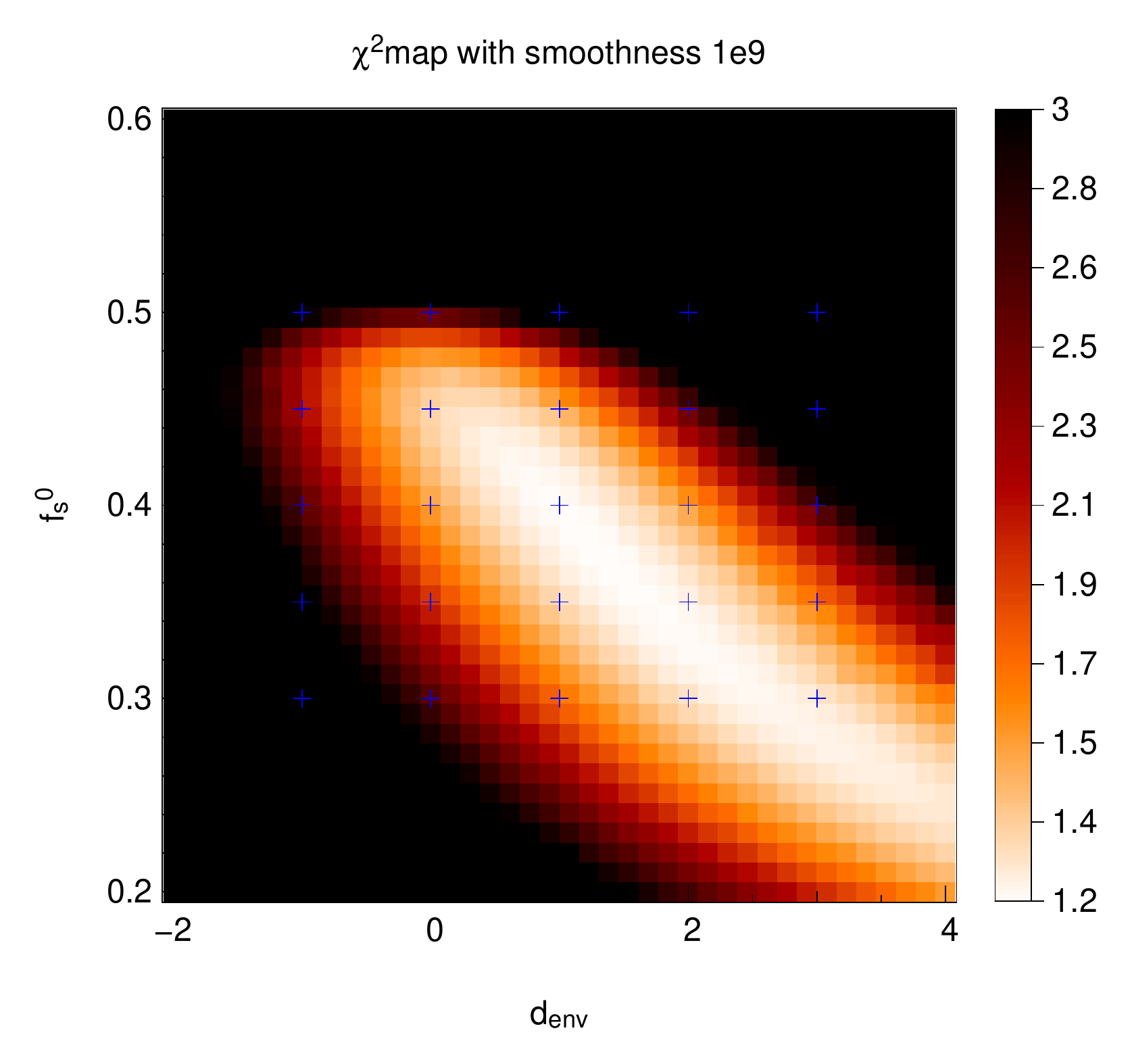}
      \includegraphics[height=4.1cm]{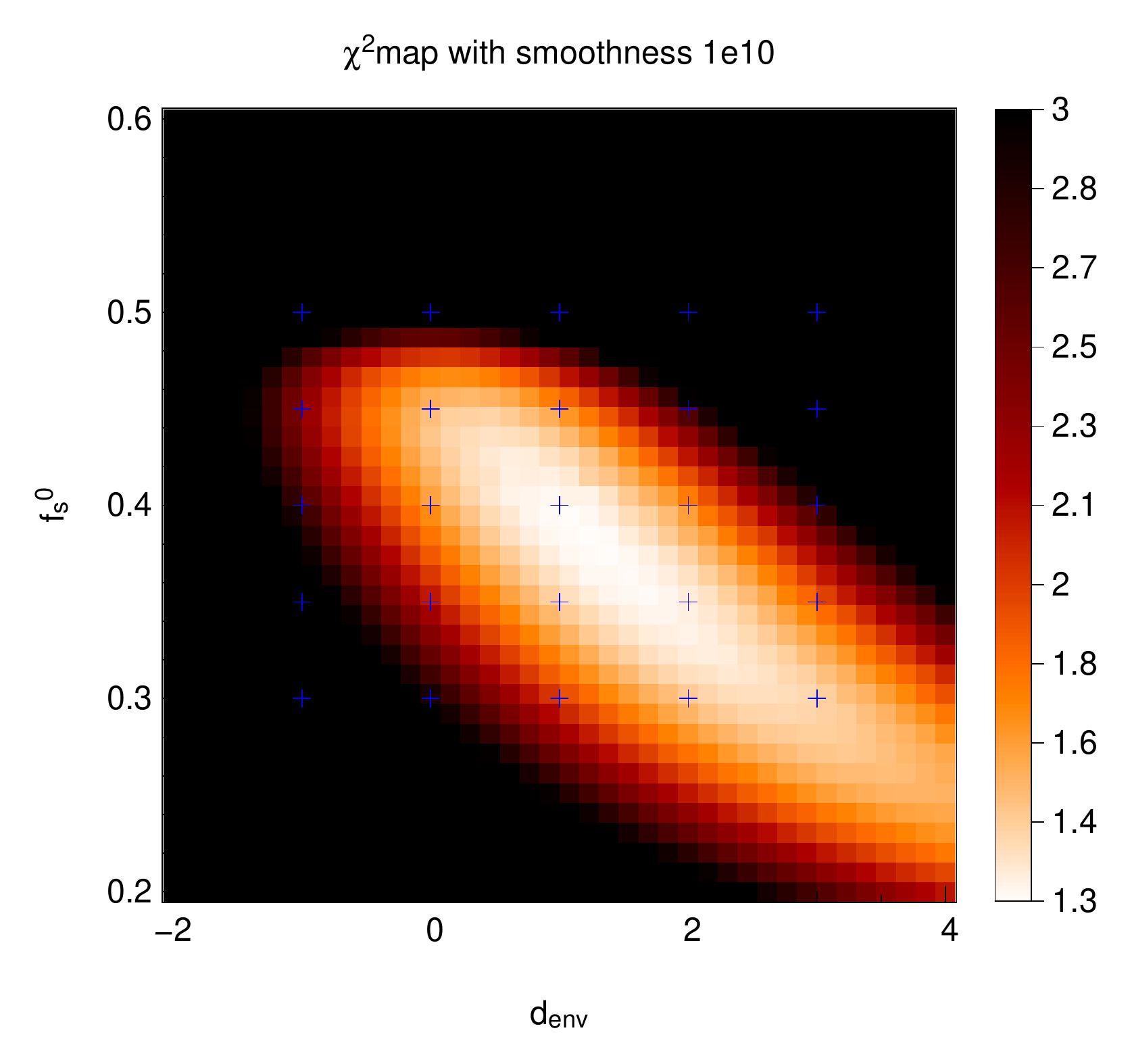}
 \caption{Choice of the $\mu$ parameter and its influence on the fit of the chromatic parameters. $\chi^2$ map of the reconstructions as a function of the chromatic parameters for reconstructions with the smoothness regularization. Black represents $\chi^2 > 3$. The blue crosses represent the location of the images of Fig.~\ref{fig:mos} in the $\chi^2$ map. The four figures represent $\mu=\{10^7, 10^8, 10^9, 10^{10}\}$ respectively. The $\chi^2$ does not change significantly with the $\mu$ hyper-parameter, but the degeneracy is always strong.}
 \label{fig:chi2sm}
\end{figure}

We can compare the $\chi^2$ map of Fig.~\ref{fig:chi2sm} realized with the smoothness regularization with Fig.~\ref{fig:mos}, which was realized with the total-variation regularization. These two maps are very similar. The degeneracy is present for both regularizations. Even when we change the hyper-parameter $\mu$, the area where the $\chi^2$values are acceptable is still large.

\section{Method of chromatic parameter fitting: }
\label{app:fit}
We propose the following algorithm iterative sequence (where $i$ is the iteration index) for a gradient-descent-based algorithm for image reconstruction (e.g. \texttt{MiRA}): 
\begin{enumerate}
\item \textit{Computing the total visibilities ($\Vtot^i$):} to compute the total complex visibilities, we use Eq.~(\ref{eqn:chrom2}) using $\Venv^i$ (which is the Fourier transform of $\mathbf{x}^i$) and the chromatic parameters ${f_*^0}^i$ and $\denv^i$ .
\item \textit{Computing the $\chi^2$-gradient:} if the modified algorithm is based on gradient minimization, it computes the flux gradient on every pixel. The first step is to compute the errors and the gradients on the total visibilities. Then the environment visibility gradient is obtained by multiplying the total visibility gradient by a factor $\frac{ \partial \Vtot} {\partial V_{\mathrm{env}}}$ (which is the environment-to-total-flux ratio). The last step is a Fourier transform of these gradients to obtain the flux gradients on the pixels.
\item \textit{Making a step in the pixel fluxes:} the algorithm will add to or subtract some flux from on every pixel depending on the gradient and the boundaries. We then have $\mathbf{x}^{i+1}$.
\item \textit{Computing the visibilities from the image:} this step is made with ${f_*^0}^i$ and $\denv^i$. We then have $\Vtot^{i+\frac{1}{2}}$.
\item \textit{Fitting the chromatic parameters:} the chromatic parameters are fit between two steps of the image reconstruction given the current image, in this way we obtain $\Vtot^{i+1}$. This step can be made using a Levenberg-Marquardt minimization. Then we return to the first step.
\end{enumerate}
where at the $i$-th iteration we have an image $\mathbf{x}^i$ and chromatic parameters ${f_*^0}^i$ and $\denv^i$.

\section{Convergence of the chromatic parameter fitting:}
\label{app:fitconv}

\begin{figure}[!t]
 \centering
      \includegraphics[height=7cm]{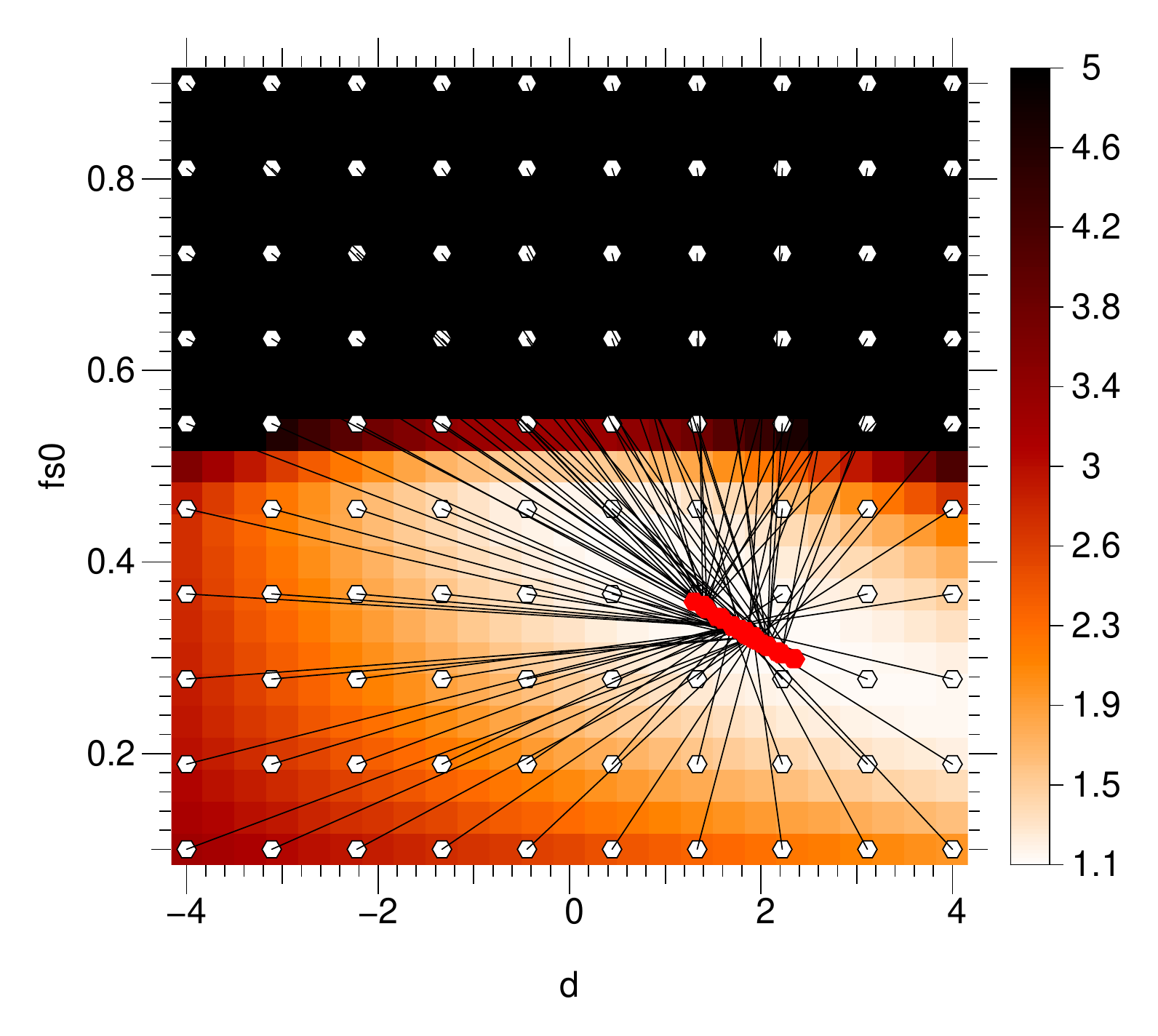}
 \caption{$\chi^2$ map with the fit departures and arrivals for the model. The color represent the $\chi^2$ for every pair of chromatic parameters $\fso$ and $\denv$. The white dots are the starting pair for image reconstructions. The red points represent the found parameters for the final image. The fit always ends in the $\chi^2$ minimum but far from the correct values ($\fso=0.4$ and $\denv=1$). The black solid lines link the start and the final values of one reconstruction.}
 \label{fig:fit}
\end{figure}

To test the results of a fit of the parameters without an SED constrain we produced a series of image reconstructions with fitting processes using a hundred of different starting points that formed a regular grid on the chromatic parameters (\fso from 10\% to 90\% and \denv from -4 to 4). The results are given in Fig \ref{fig:fit}. The results of all the fitting processes end in the $\chi^2$ valley, but not always close to the good values. This underlines that it is important to know the spectro-photometric information of the observed target and that the degeneracy is deeply embedded in the interferometric data.

To conclude, the fitting process finds the global minimum valley. The starting points do not influence the convergence (see Fig~\ref{fig:fit}).

\end{document}